\pdfoutput=1

\documentclass[11pt,twoside,a4paper,cmspaper,final,collab]{cms-tdr}
\def\svnVersion{199416}\def\cmsCernNo{2012-252}\def\cmsCernDate{\today}\def\cmsMessage{Published in the European Physical Journal C as \href{http://dx.doi.org/10.1140/epjc/s10052-013-2493-8}{\doi{10.1140/epjc/s10052-013-2493-8.}}}

\begin{document}\cmsNoteHeader{SUS-12-004}

\hyphenation{had-ron-i-za-tion}
\hyphenation{cal-or-i-me-ter}
\hyphenation{de-vices}

\RCS$Revision: 192158 $
\RCS$HeadURL: svn+ssh://svn.cern.ch/reps/tdr2/papers/SUS-12-004/trunk/SUS-12-004.tex $
\RCS$Id: SUS-12-004.tex 192158 2013-06-24 14:41:27Z alverson $
\providecommand{\PSGt}{\ensuremath{\widetilde{\Pgt}\xspace}}
\newcommand{\WW}{\ensuremath{\PWp\PWm}\xspace}
\newcommand{\tauh}{\ensuremath{\tau_\mathrm{h}}\xspace}

\newcommand{\RTPreSelWJets}{\ensuremath{452}}
\newcommand{\RTPreSelWJetsStatUncert}{\ensuremath{30}}
\newcommand{\RTPrePredWJets}{\ensuremath{441}}
\newcommand{\RTPrePredWJetsStatUncert}{\ensuremath{21}}
\newcommand{\RTFullSelWJets}{\ensuremath{28.9}}
\newcommand{\RTFullSelWJetsStatUncert}{\ensuremath{7.5}}
\newcommand{\RTPreSelTTbar}{\ensuremath{60.6}}
\newcommand{\RTPreSelTTbarStatUncert}{\ensuremath{3.7}}
\newcommand{\RTFullPredWJets}{\ensuremath{34.9}}
\newcommand{\RTFullPredWJetsStatUncert}{\ensuremath{5.9}}
\newcommand{\RTPrePredTTbar}{\ensuremath{63.2}}
\newcommand{\RTPrePredTTbarStatUncert}{\ensuremath{2.1}}
\newcommand{\RTFullSelTTbar}{\ensuremath{1.6}}
\newcommand{\RTFullSelTTbarStatUncert}{\ensuremath{0.6}}
\newcommand{\RTFullPredTTbar}{\ensuremath{2.9}}
\newcommand{\RTFullPredTTbarStatUncert}{\ensuremath{0.4}}
\newcommand{\RTPreSelZJets}{\ensuremath{10.9}}
\newcommand{\RTPreSelZJetsStatUncert}{\ensuremath{2.1}}
\newcommand{\RTPrePredZJets}{\ensuremath{8.4}}
\newcommand{\RTPrePredZJetsStatUncert}{\ensuremath{1.3}}
\newcommand{\RTFullSelZJets}{\ensuremath{0.8}}
\newcommand{\RTFullSelZJetsStatUncert}{\ensuremath{0.6}}
\newcommand{\RTFullPredZJets}{\ensuremath{0.4}}
\newcommand{\RTFullPredZJetsStatUncert}{\ensuremath{0.3}}
\newcommand{\RTPreSelWWJets}{\ensuremath{15.1}}
\newcommand{\RTPreSelWWJetsStatUncert}{\ensuremath{1.6}}
\newcommand{\RTPrePredWWJets}{\ensuremath{14.4}}
\newcommand{\RTPrePredWWJetsStatUncert}{\ensuremath{1.1}}
\newcommand{\RTFullSelWWJets}{\ensuremath{0.5}}
\newcommand{\RTFullSelWWJetsStatUncert}{\ensuremath{0.3}}
\newcommand{\RTFullPredWWJets}{\ensuremath{1.3}}
\newcommand{\RTFullPredWWJetsStatUncert}{\ensuremath{0.3}}
\newcommand{\RTPreSelAll}{\ensuremath{539}}
\newcommand{\RTPreSelAllStatUncert}{\ensuremath{30}}
\newcommand{\RTFullPredAll}{\ensuremath{39.5}}
\newcommand{\RTFullPredAllStatUncert}{\ensuremath{5.9}}
\newcommand{\RTFullSelAll}{\ensuremath{31.8}}
\newcommand{\RTFullSelAllStatUncert}{\ensuremath{7.5}}
\newcommand{\RTPrePredAll}{\ensuremath{527}}
\newcommand{\RTPrePredAllStatUncert}{\ensuremath{21}}

\newlength\cmsFigWidth
\ifthenelse{\boolean{cms@external}}{\setlength\cmsFigWidth{0.85\columnwidth}}{\setlength\cmsFigWidth{0.4\textwidth}}
\ifthenelse{\boolean{cms@external}}{\providecommand{\cmsLeft}{top}}{\providecommand{\cmsLeft}{left}}
\ifthenelse{\boolean{cms@external}}{\providecommand{\cmsRight}{bottom}}{\providecommand{\cmsRight}{right}}
\ifthenelse{\boolean{cms@external}}{\titlerunning{search for BSM physics with $\tau$s, jets, and large \pt}}{}
\ifthenelse{\boolean{cms@external}}{\providecommand{\MHT} {\ensuremath{H_\mathrm{T}\hspace{-1.3em}/\kern0.7em}\xspace}}{\newcommand{\MHT} {\ensuremath{H_\mathrm{T}\hspace{-1.2em}/\kern0.55em}\xspace}}
\cmsNoteHeader{SUS-12-004} 
\title{Search for physics beyond the standard model in events with $\tau$ leptons, jets, and large transverse momentum imbalance in pp collisions at $\sqrt{s}=7$\TeV}

\date{\today}

\abstract{
A search for physics beyond the standard model is performed with events having
one or more hadronically decaying $\tau$ leptons, highly energetic jets, and large transverse momentum imbalance.
The data sample corresponds to an integrated luminosity of 4.98\fbinv of proton-proton collisions at $\sqrt{s}=7$\TeV collected with the CMS detector at the
LHC in 2011.
The number of observed events is consistent with predictions for standard model processes.
Lower limits on the mass of the gluino in supersymmetric models are determined.
}

\hypersetup{%
pdfauthor={CMS Collaboration},%
pdftitle={Search for physics beyond the standard model in events with tau leptons, jets, and large transverse momentum imbalance in pp collisions at sqrt(s)=7 TeV},%
pdfsubject={CMS},%
pdfkeywords={CMS, physics, supersymmetry}}

\maketitle 

\section{Introduction}\label{sec:intro}

The standard model (SM) of particle physics has been successful
in explaining a wide variety of data.
In spite of this, the SM is incomplete.
For example, it possesses a divergence in the Higgs sector \cite{HiggsH}
and has no cold dark matter (DM) candidate \cite{WMAP}.
Many models of physics beyond the SM (BSM) have been proposed in order to address these problems.

DM particles, if produced in proton-proton collisions at the CERN Large Hadron Collider (LHC), would escape
detection and result in a significant transverse momentum ($\pt$) imbalance in the detector.
Additionally, cascade decays of heavy colored particles
to final states with a high multiplicity of energetic jets and $\tau$ leptons appear very naturally in many BSM physics scenarios.
Hence, events with multiple $\tau$ lepton candidates, large jet multiplicity, and significant transverse momentum imbalance,
represent a distinct signature of new physics.
In this paper, focus is placed on final states with hadronically decaying $\tau$ leptons.
In what follows, the visible part of a hadronically decaying $\tau$ lepton will be referred to as $\tauh$.

In certain models of supersymmetry (SUSY), the lightest supersymmetric particle (LSP) is a candidate
for DM. It has been appreciated for some time that the DM relic density
may be sensitive to coannihilation processes involving the LSP and the next-to-lightest
supersymmetric particle (NLSP).
Coannihilation is characterized by a mass difference ($\Delta M$)
between the NLSP and the LSP of approximately 5--15\GeV \cite{Martin:1997ns, Wess:1974tw,ArnowittSUGRAGUT, HallSUGRAGUT}.
This small mass difference would be
necessary to allow the NLSP to coannihilate with the LSP in the early universe, leading to the
dark matter abundance that is currently observed \cite{Griest}.
If the supersymmetric partner of the $\tau$ lepton, the stau ($\PSGt$), is the NLSP, and if the $\PSGt$ decays
primarily to a $\tau$ lepton and the LSP, small values of $\Delta M$ would lead to final states with low-energy $\tau$ leptons
($\pt \sim \Delta M$) \cite{CoannihilationPaper}.
Decays of colored SUSY particles can produce the $\PSGt$ via chargino ($\PSGc^{\pm}$) or neutralino ($\PSGcz$) intermediate states
(e.g., $\PSGczDt\to\tau\PSGt\to\tau\tau\PSGczDo$), resulting in final states with at least one $\tauh$.

We present a search for BSM particles in events with exactly
one $\tauh$ lepton and jets (single-$\tauh$ final state), and in events with jets and two or more $\tauh$ leptons (multiple-$\tauh$ final state).
These two topologies provide complementary sensitivity to models with a wide range of $\Delta M$ values.
For example, in the case of very small values of $\Delta M$ ($\sim$5\GeV), the low-energy
$\tauh$ cannot be effectively detected and the search for new physics in the single-$\tauh$ final state has better sensitivity.
The analysis is performed using proton-proton collision data at $\sqrt{s} = 7$\TeV collected with the Compact Muon Solenoid (CMS) detector
\cite{CMS} at the LHC in 2011.
The data sample corresponds to an integrated luminosity of $4.98 \pm0.11\fbinv$.
The search is characterized by methods that determine the backgrounds directly from data, to reduce the reliance on simulation.
To illustrate the sensitivity of this search to BSM processes, the constrained minimal
supersymmetric extension of the standard model, or minimal supergravity,
is chosen as the benchmark \cite{Martin:1997ns, CMSSM, MSUGRA}; we denote this benchmark as ``CMSSM".
An interpretation of the results in the context of simplified model spectra (SMS)
\cite{SMS1,SMS2} is also presented.
The ATLAS collaboration has published a result on a search for one or more hadronically
decaying tau leptons, highly energetic jets, and a large transverse momentum imbalance
probing minimal Gauge Mediated Symmetry Breaking (GMSB) models ~\cite{AtlasTauPaper}.

\section{The CMS detector}
The central feature of the CMS apparatus is a
superconducting solenoid, of 6\unit{m} inner diameter, providing a magnetic field
of 3.8\unit{T}. Within the field volume are a silicon pixel and strip
tracker, a crystal electromagnetic calorimeter (ECAL),
which includes a silicon sensor preshower detector in front of the ECAL endcaps,
and a brass-scintillator hadron calorimeter. Muons are measured in
gas-ionization detectors embedded in the steel return yoke. In
addition to the barrel and endcap detectors, CMS has extensive forward
calorimetry.

The inner tracker measures charged particles within $|\eta| < 2.5$
and provides an impact parameter resolution of about 15 $\mu$m and a $\pt$ resolution of about 1.5\% for
100\GeV particles.
Collision events are selected with a first-level trigger based on fast electronics, and a higher-level trigger that
runs a version of the offline reconstruction program optimized for speed.

The CMS  experiment
uses a right-handed coordinate system, with the origin at the
nominal interaction point, the $x$ axis pointing to the center of
the LHC ring, the $y$ axis pointing up (perpendicular to the plane of the LHC ring),
and the $z$ axis along the counterclockwise beam direction. The polar
angle $\theta$ is measured from the positive $z$ axis and the
azimuthal angle in the $x$-$y$ plane.
The pseudorapidity is given by $\eta = -\ln[\tan(\theta/2)]$.

\section{Object reconstruction and identification}\label{sec:leptonRecoId}

Jets in the detector are reconstructed using particle-flow (PF) objects~\cite{CMS-PAS-PFT-10-002}.
In the PF approach, information from all subdetectors is combined to reconstruct and
identify final-state particles (muons, electrons, photons, and charged and neutral hadrons) produced in the collision.
The anti-\kt clustering algorithm~\cite{antikt} with a distance parameter $R  =  0.5$
is used for jet clustering.
Jets are required to satisfy criteria designed to identify anomalous behavior in the calorimeters, and to be well
separated from any identified $\tau$ lepton.

Validation and efficiency studies are performed utilizing events with a $\tauh$
lepton and a light-lepton $\ell$, with $\ell$ representing an electron (\Pe) or muon (\Pgm).
Muons are reconstructed using the tracker and muon chambers.
Selection requirements based on the minimum number of hits in the silicon
tracker, pixel detector, and muon chambers are applied to suppress
muon backgrounds from decays-in-flight of pions or kaons \cite{MUONreco}.
Electrons are reconstructed by combining tracks with ECAL clusters.
Requirements are imposed to distinguish between prompt and non-prompt electrons, where the
latter can arise from charged pion decay or photon conversion \cite{ELECTRONreco}.
The light-lepton candidates are required to satisfy both track and ECAL isolation requirements.
The track isolation variable is defined as the sum of the \pt\ of the tracks, as measured by
the tracking system, within an isolation cone of radius $\Delta R = \sqrt{(\Delta\eta)^{2} + (\Delta\phi)^{2}}=0.4$ centered on
the light-lepton track. The ECAL isolation variable is based on the amount of energy deposited in the ECAL within the same isolation cone.
In both cases the contribution from the light-lepton candidate is removed from the sum.

Reconstruction of hadronically decaying $\tau$ leptons is performed using the hadron-plus-strips (HPS) algorithm \cite{TauPAS},
designed to optimize the performance of $\tauh$ reconstruction by considering specific $\tauh$
decay modes. To suppress backgrounds in which light-quark or gluon jets mimic hadronic $\tau$ decays, a $\tauh$ candidate is
required to be spatially isolated from other energy deposits in the calorimeter. Charged hadrons and photons not considered
in the reconstruction of the $\tauh$ decay mode are used to calculate the isolation.
Additionally, $\tauh$ candidates are required to be distinguished from electrons and muons in the event.
In this analysis, two HPS isolation definitions are used.
The $\tauh$ isolation definition used for single-$\tauh$ final states rejects a $\tauh$ candidate if one or more charged hadrons
with $\pt > 1.0$\GeV or one or more photons with transverse energy $\ET > 1.5$\GeV is found within an isolation cone of radius $\Delta R = 0.5$.
The $\tauh$ isolation definition used for multiple-$\tauh$ final states rejects a $\tauh$ candidate if one or more charged hadrons
with $\pt > 1.5\GeV$ or one or more photons with transverse energy $\ET > 2.0$\GeV is found within an isolation cone of radius $\Delta R = 0.3$.
The isolation criteria used for the multiple-$\tauh$ final state increases the signal-to-background ratio while reducing the rate of
$\tauh$ misidentification. This affects the yield of events with light-quark or gluon jets that are misidentified as $\tauh$ leptons,
which depends on the square of the misidentification rate.
Here a final state with exactly two $\tauh$ candidates is considered since events with more than two  $\tauh$ candidates are only
a small fraction ($<$1\%) of events.

The missing transverse momentum $\MHT$ is defined as:
\begin{equation}
 {\MHT} = \left| \sum \ptvec^{\text{jet}} \right|,
\label{eq:MHT}
\end{equation}
where the sum runs over all the jets with $\pt^\text{jet} > 30$\GeV inside the fiducial detector volume of $|\eta| < 5$. The vector $\vec{\MHT}$
 is the negative of the vector sum in Eq.~(\ref{eq:MHT}).
The observable $\HT = \sum \ptvec^{\text{jet}}$ is used to estimate the overall energy scale of the event.
For the single-$\tauh$ final state, \HT\ is calculated using jets with $\pt > 50$\GeV
and will be referred to as $\HT^{50}$.
For the multiple-\tauh\ final state, \HT\ is calculated using jets with $\pt > 30$\GeV
and will be referred to as $\HT^{30}$.
In both instances of the \HT\ calculation, we consider all jets in $|\eta| < 5 $ (the fiducial detector limit).
The use of a lower \pt\ threshold for the jets in the multiple-\tauh\ final state
increases the efficiency of signal events without
significantly increasing the background.
\section{Signal and background samples}\label{sec:backgrounds}

The major sources of SM background are top-quark pair (\ttbar) events and events with a W or Z boson accompanied by jets.
Both \ttbar and $\PW+\text{jets}$ events can have genuine $\tauh$ leptons, large genuine \MHT\ from
W boson decays, and jets that can be misidentified as a $\tauh$.
Similarly, $\cPZ + \text{jets}$  events with $\cPZ (\to \nu\nu)$ and with one or more jets misidentified as a $\tauh$ lepton provide a source
of background. $\cPZ +  \text{jets}$ events with $\cPZ (\to \nu\nu)$ present a background because of the genuine $\tauh$ leptons and the genuine
\MHT\ from the neutrinos in the $\tauh$ decay.
QCD multijet events can become a background when a mismeasured jet gives rise to large
\MHT\ and jets are misidentified as $\tauh$ leptons.

Data are compared with predictions obtained from samples of Monte Carlo (MC) simulated events. Signal and background MC
samples are produced with the \PYTHIA 6.4.22\,\cite{PYTHIA} and \MADGRAPH\,\cite{Madgraph}
generators using the Z2 tune \cite{TuneZ2}
and the NLO CTEQ6L1 parton distribution function (PDF)
set \cite{CTEQL1}.
The $\tau$ lepton decays are simulated with the \textsc{Tauola} \cite{TAUOLA} program.
The generated events are processed with a detailed simulation of the CMS apparatus using
the {\GEANTfour} package \cite{Geant}. The MC yields are normalized to the integrated luminosity of the data using
next-to-leading order (NLO) cross sections \cite{ref:CMSSMuncerts, ref:CMSSMuncerts2, ref:CMSSMuncerts3, ref:CMSSMuncerts4, ref:CMSSMuncerts5, ref:CMSSMuncerts6}.
For the 2011 LHC running conditions, the mean number of
interactions in a single beam crossing is $\sim$10.
The effect of multiple interactions per bunch crossing (pileup) is taken into account by superimposing MC minimum-bias events so that the probability
distribution for overlapping pp collisions in the simulation matches the measured distribution.

\section{Event selection}\label{sec:data}

Events for both the single- and multiple-$\tauh$ final states are selected using a trigger that requires $\MHT > 150$\GeV.
This trigger allows us to maintain sensitivity in regions where the \pt\ value of the \tauh\ is small ($\pt \sim 15$\GeV).
This trigger efficiency, for an offline selection requirement of $\MHT > 250$\GeV, is $98.9\%$.
For the $\tauh$ efficiency and validation studies, samples are chosen using triggers that require the presence of both
a \tauh\ candidate and a muon.

The \tauh\ candidates must satisfy $\pt >15$\GeV and $|\eta|<2.1$. For the single-\tauh\ final state we require that no additional light leptons
be present in the event. This requirement suppresses background from $\ttbar$, $\PW+\text{jets}$, and Z + jets events. For the multiple-$\tauh$
final state there is no requirement placed on the number of light leptons.

For the single-\tauh\ final state, we define a baseline event selection $\HT^{50} > 350$\GeV and $\MHT>250$\GeV. The sample obtained with
the baseline selection is used to validate the background predictions. The signal region (SR) for the single-\tauh\ final state is defined by
$\HT^{50} > 600$\GeV and $\MHT>400$\GeV.

For the multiple-$\tauh$ final state, 
the SR is defined by ${\MHT} > 250$\GeV and by the requirement that there be at least two jets
with $\pt > 100$\GeV and $|\eta| < 3.0$.
QCD multijet events are rejected by requiring the azimuthal difference
$\Delta \phi (j_{2},\MHT)$
between the second leading jet in \pt and $\vec{\MHT}$ to satisfy
$|\Delta \phi (j_{2},\MHT)|>0.5$
Finally, events are required to contain at least one $\tauh\tauh$ pair
separated by $\Delta R ({\tauh}_{,i},{\tauh}_{,j})>0.3$.

\section{Background estimate}\label{sec:singletaus}

The background contributions are categorized differently for the single- and multiple-$\tauh$ final states.
For the single-\tauh\ final state,
the background contributions are divided into events containing a genuine \tauh\
and events where a jet is misidentified as a \tauh.
For the multiple-\tauh\ final state, the main background contribution arises from misidentified \tauh\ leptons.
We identify the different sources of background individually using dedicated data control regions (CR).

\subsection{Estimate of backgrounds in the single-\texorpdfstring{\tauh}{tau(h)} final state}

In the single-\tauh\ final state, the largest background contribution comes from W $+$ jets events that contain a genuine \tauh\ lepton.
The other significant contribution arises from QCD multijet events in which a jet is misidentified as a \tauh.
The $\PW+\text{jets}$ background contribution is estimated using a sample of $\PW+\text{jets}$ events with $\PW\to\mu\nu$. The QCD multijet background is
determined by selecting a QCD-dominated CR and evaluating the $\tauh$ misidentification rate.

\subsubsection{Estimate of the W + jets background in the single-\texorpdfstring{\tauh}{tau(h)} final state}

To evaluate the $\PW+\text{jets}$ background, we exploit the similarity between W decays to a muon and to a tau lepton and select
a sample of $\PW+\text{jets}$ events with $\PW(\to \mu\nu$). This sample will be referred to as the muon control sample.
To select the muon control sample, events are required  to contain exactly one muon and no reconstructed $\tauh$
or electron. To emulate the $\tauh$ acceptance, the muon is
required to satisfy $|\eta| < 2.1$.
The yields in the muon control sample are corrected for muon reconstruction 
($\varepsilon_{\mu}^{\text{reco}}$) and isolation efficiency ($\varepsilon_{\mu}^{\text{iso}}$).
The muon reconstruction efficiency is derived from data using a sample of Z + jets
events and parameterized as a function of $\pt$ and $\eta$. 
The muon isolation criteria
help to distinguish between muons from the decay of the W boson and muons
from semileptonic decays of c and b quarks.
The isolation efficiency is parameterized as a function of
the separation from the nearest jet and the momentum of the jet.
A correction factor ($P^\PW_{\mu}$) is applied to the muons in the muon control sample to account for
muons that do not come from a $\tau$-lepton decay. This correction factor
depends on the $\pt$ of the muon and the \MHT\  value in the event and is derived
from a simulated sample of $\PW+\text{jets}$ events.

\begin{figure}[htb]
\begin{center}
\includegraphics[width=.45\textwidth]{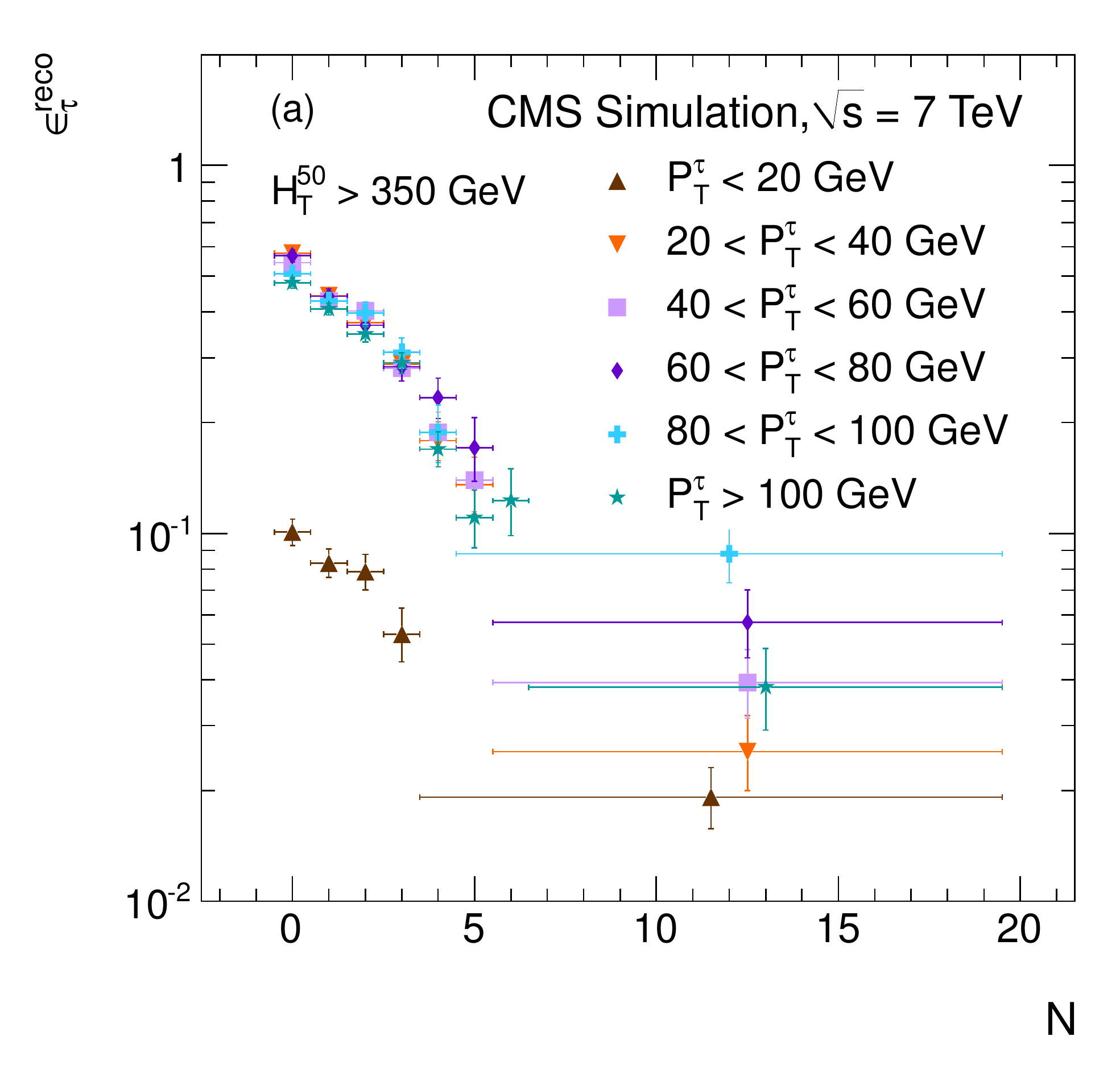}
\includegraphics[width=.45\textwidth]{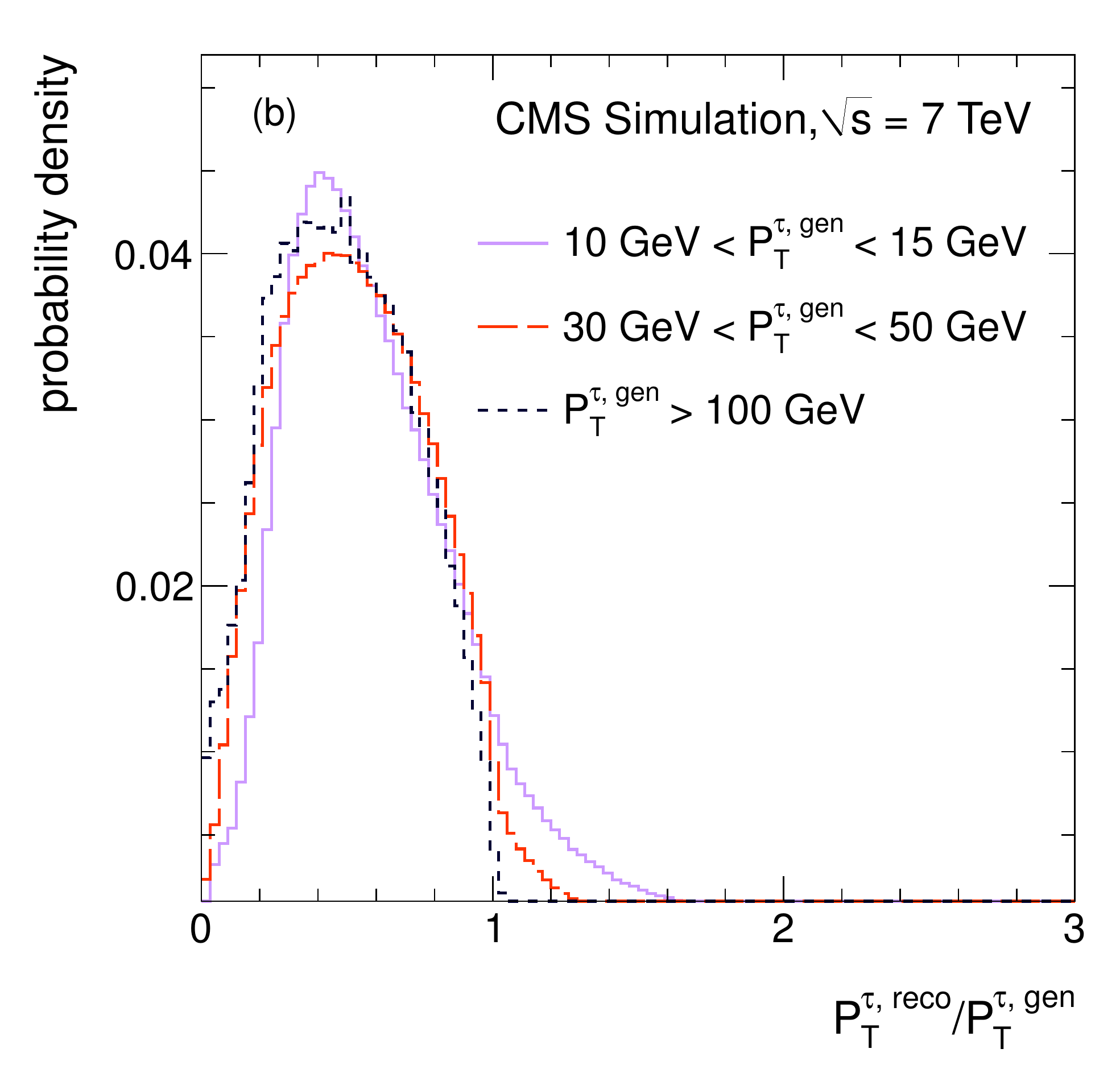}
\end{center}
\caption{(a) Dependence of the $\tauh$ reconstruction efficiency $\epsilon_{\tau}^{\text{reco}}$ on the number of additional particles $N$ in the
isolation cone in bins of $\tauh$ lepton $\pt$ for the single-\tauh\ final state, where $N$ is the total number of the photons and charged hadrons in the
isolation cone, and (b) dependence of $\tauh$ response on $\PT^{\tau, \text{gen}}$.
Both distributions are derived from a simulated sample of
$\PW(\to \tau\nu) + \text{jets}$ events.}
\label{fig:NotFromTau}
\end{figure}

As the muons in the muon control sample are selected to mimic a $\tauh$,
a correction is applied to emulate the probability to reconstruct and identify a $\tauh$ lepton.
The reconstruction and identification efficiency $\varepsilon_{\tau}^{\text{reco}}$ is parameterized as a function of the $\pt$
of the \tauh\ candidate and as a function of
the total number $N$ of charged particles and photons in the isolation cone [Fig.~\ref{fig:NotFromTau}(a)].
Corrections are also applied to account for the hadronic branching fraction ($f_{\tau}^{\text{bf(hadr)}}$) of a $\tau$ lepton.
Except for the $f_{\tau}^{\text{bf(hadr)}}$ 
the values of the correction factors differ in each event.
The corrections are combined to define an overall event weight, defined as:
\begin{equation}
  f^\text{corr}_\text{event} = \frac{P_{\mu}^{\PW} \times \varepsilon_{\tau} \times f_{\tau}^{\text{bf(hadr)}}}{\varepsilon_{\mu}^{\text{reco}}
\times \varepsilon_{\mu}^{\text{iso}}}.
\end{equation}

A $\tauh$ response template is derived from simulated events. The response template is given by the ratio of the reconstructed energy of the
$\tauh$ to the true generator-level energy. The $\tauh$ response depends on the transverse momentum
of the generated $\tau$ lepton [Fig.~\ref{fig:NotFromTau}(b)] and on the number of reconstructed primary vertices in the event. The muon
\PT\ spectrum is smeared as a function of \PT\ and the number of primary vertices to mimic the \PT\ distribution of the $\tauh$.

\begin{figure}[htb]
\begin{center}
\includegraphics[width=.45\textwidth]{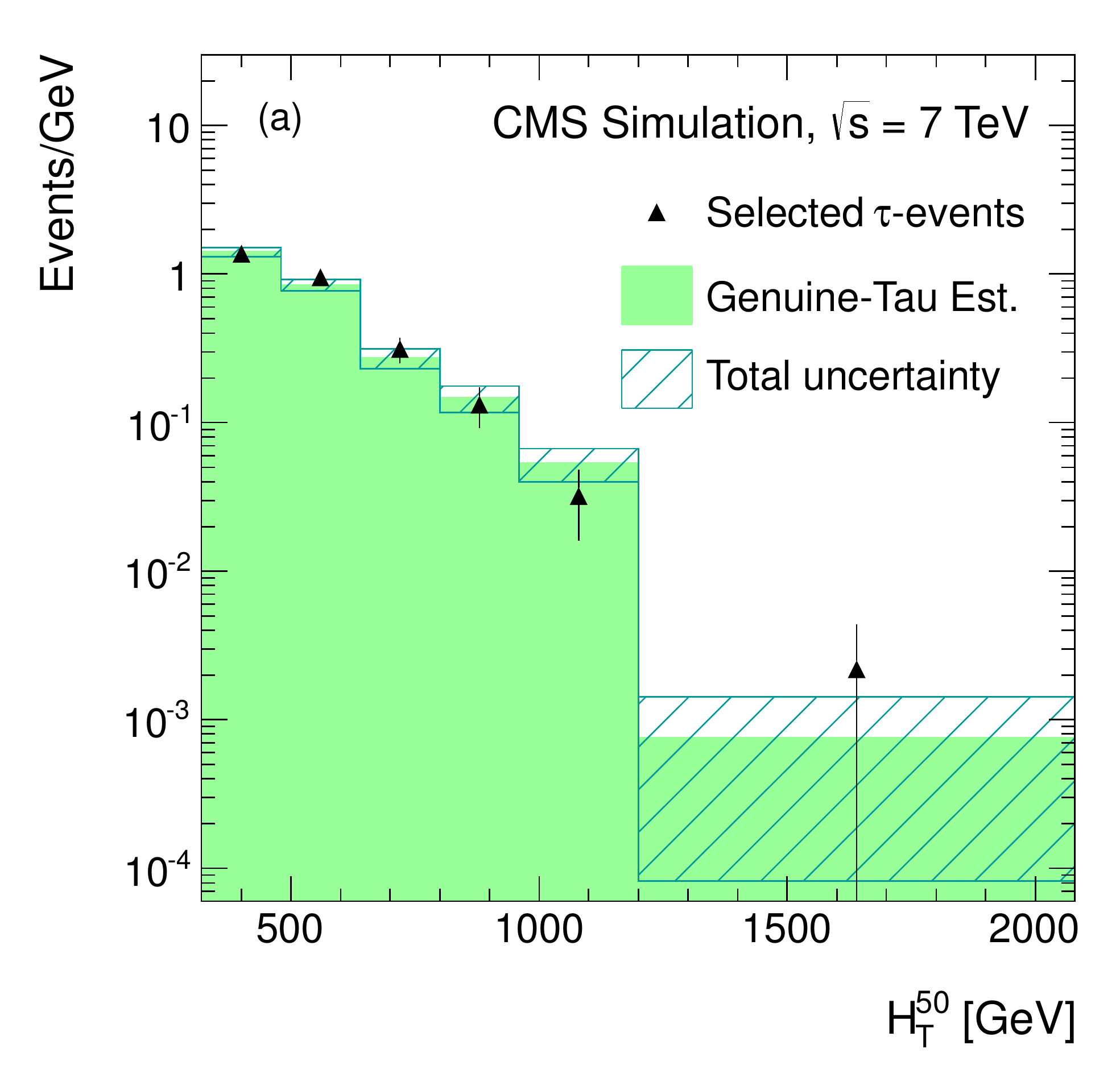}
\includegraphics[width=.45\textwidth]{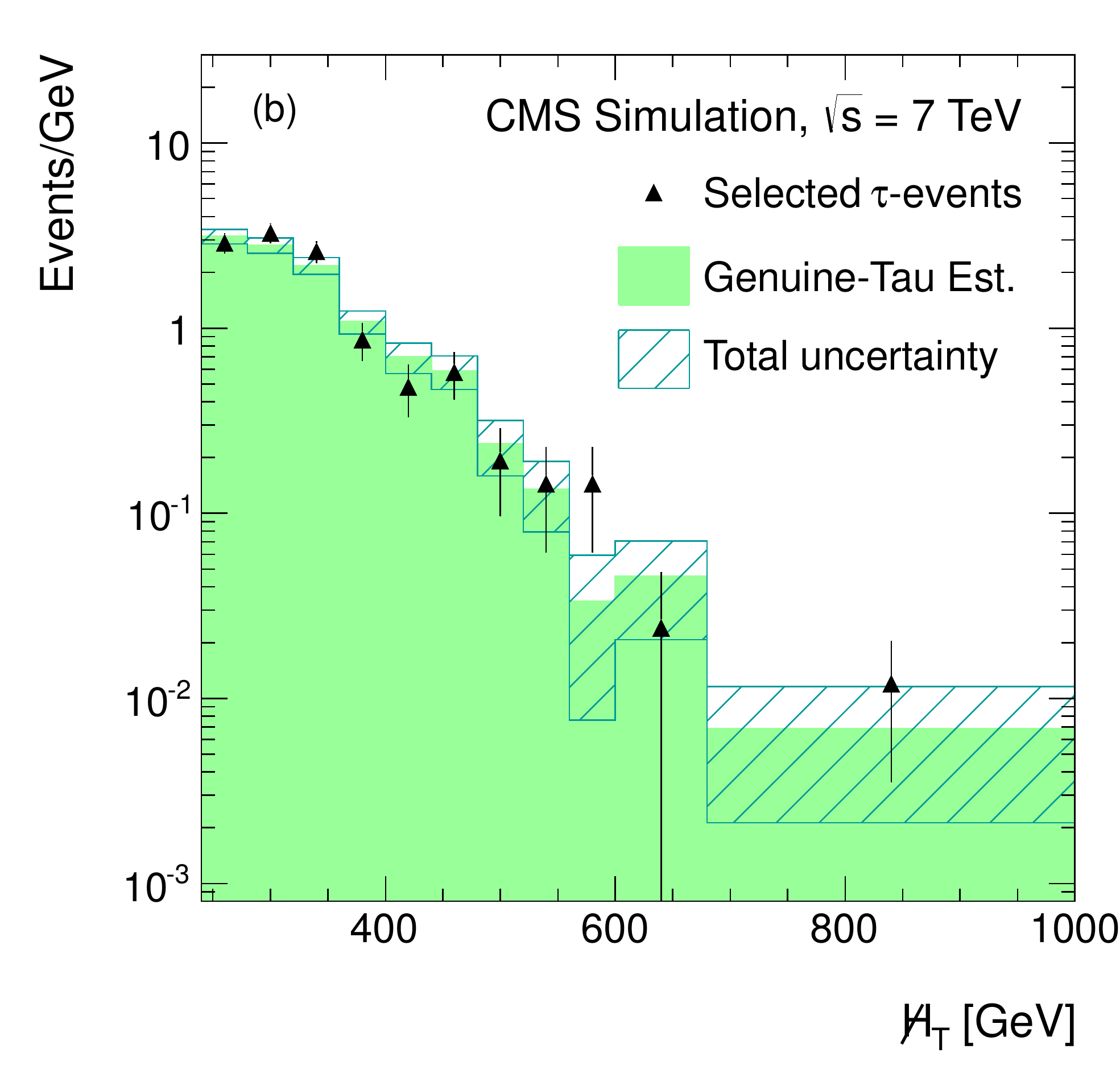}
\end{center}
\caption{Distributions of (a) $\HT^{50}$ and (b) \MHT\ for the genuine $\tauh$ estimate in simulated $\PW+\text{jets}$ events for the single-\tauh\
final state. The black triangles show the results for events that satisfy the baseline selection and that contain a reconstructed
$\tauh$ matched to the visible part of a
generated, hadronically decaying $\tau$ lepton. The filled green areas show the prediction obtained from the simulated muon control sample. The hatched areas are the total
uncertainty on the prediction. }
 \label{fig::realTauMCClosureWJetsControl1}
\end{figure}

Fully simulated $\PW + \text{jets}$ events are used to verify the procedure.
Figure~\ref{fig::realTauMCClosureWJetsControl1} shows the $\HT^{50}$\ and \MHT\ distributions from simulated $\PW+\text{jets}$ events
for the single-$\tauh$ final state. These events satisfy the baseline selection described in Section \ref{sec:data}.
The reconstructed $\tauh$ is required to match a hadronically decaying generated tau lepton, to ensure that only the
genuine tau background is addressed in this check.
The event yield and distributions are compared with the prediction from the simulated muon control sample and agree within statistical
uncertainties, thus verifying the closure of the method in MC simulation.
Hence, the predicted $\HT^{50}$\ and \MHT\ distributions from the muon control sample can be taken to describe a $\tauh$
sample within statistical uncertainties.

\begin{table*}[tbh]
\begin{center}
\topcaption{The selected and predicted background contributions for simulated events with a genuine $\tauh$ passing the baseline
and signal selection in the single-\tauh\ final state. The reconstructed $\tauh$ is required to match the visible part of the generated, hadronically
decaying $\tau$-lepton. The predictions are derived from the muon control sample.
  }
\label{tab::allMCRealTauBaseLine}
\begin{tabular}{ccccc}
\hline
\multirow{2}{*}{L = 4.98\fbinv}&\multicolumn{2}{c}{Baseline Selection}&\multicolumn{2}{c}{Signal Selection}\\
& Selected & Predicted & Selected & Predicted \\ \hline
${\PW}(\to \ell\nu) +\text{jets}$ & $\RTPreSelWJets \pm \RTPreSelWJetsStatUncert$ & $\RTPrePredWJets\pm \RTPrePredWJetsStatUncert$ & $\RTFullSelWJets \pm \RTFullSelWJetsStatUncert$ & $\RTFullPredWJets\pm \RTFullPredWJetsStatUncert$ \\
\ttbar  & $\RTPreSelTTbar \pm \RTPreSelTTbarStatUncert$ & $\RTPrePredTTbar\pm \RTPrePredTTbarStatUncert$ & $\RTFullSelTTbar \pm \RTFullSelTTbarStatUncert$ &
 $\RTFullPredTTbar\pm \RTFullPredTTbarStatUncert$ \\
$\cPZ(\to \ell\ell) +$ jets & $\RTPreSelZJets \pm \RTPreSelZJetsStatUncert$ & $\RTPrePredZJets \pm \RTPrePredZJetsStatUncert$ & $\RTFullSelZJets \pm \RTFullSelZJetsStatUncert$ & $\RTFullPredZJets\pm \RTFullPredZJetsStatUncert$ \\
\WW & $\RTPreSelWWJets \pm \RTPreSelWWJetsStatUncert$ & $\RTPrePredWWJets\pm \RTPrePredWWJetsStatUncert$ & $\RTFullSelWWJets \pm \RTFullSelWWJetsStatUncert$ & $\RTFullPredWWJets\pm \RTFullPredWWJetsStatUncert$ \\
Sum  & $\RTPreSelAll \pm \RTPreSelAllStatUncert$ & $\RTPrePredAll\pm \RTPrePredAllStatUncert$ & $\RTFullSelAll \pm \RTFullSelAllStatUncert$ & $\RTFullPredAll\pm \RTFullPredAllStatUncert$ \\ \hline
\end{tabular}
\end{center}
\end{table*}

The muon control sample consists primarily of $\PW+\text{jets}$ events, but also contains $\ttbar$ events in which one W boson decays into a muon while the other $\PW$ boson decays
either into an unidentified $\tau$ lepton or into a light lepton that is not reconstructed.
Any isolated muons produced through the decay of b or c quarks can also contribute to the 
muon control sample. SM processes containing a Z boson or two $\PW$ bosons
can also contribute to the muon control sample if one of the two decay muons is not reconstructed.

The true event yields of each process as determined from simulation
are summarized in Table~\ref{tab::allMCRealTauBaseLine} for the baseline and SR selections.
For both selections the number of predicted events with a genuine $\tauh$ lepton is seen to agree with the true number of events.
The value of $\varepsilon_{\tau}^{\text{reco}}$ that is used to calculate the predicted rate is measured in a sample of $\PW+\text{jets}$
and is different from the value that would be measured in a sample of \ttbar events.
This leads to an overestimation of the \ttbar\ contribution. A systematic uncertainty is assigned
to account for this overestimation.

\subsubsection{Estimate of the QCD multijet background in the single-\texorpdfstring{\tauh}{tau(h)} final state}

To estimate the background where a jet is misidentified as a $\tauh$ lepton, a QCD-dominated control sample is obtained by selecting events with $\HT^{50} > 350$\GeV and $40 < {\MHT} < 60$\GeV.
The control sample is selected using a prescaled $\HT$ trigger with criteria
that lead to a sample where about 99\% of the events arise from QCD multijet production.
The probability for a jet to be misidentified as a $\tauh$ lepton is measured by determining the fraction of jets from the single-$\tauh$ control sample
that pass the $\tauh$ identification criteria.
Jets considered in the calculation of the misidentification rate satisfy the requirements $\pt>5$\GeV and $|\eta|<2.5$.
The misidentification rates $f_{i}$ for each jet $i$ depend on $\eta$ and $\pt$ and are used to determine an overall weight, which is applied to each event.
The event weights are defined as:
\begin{figure}[htb]
\begin{center}
\includegraphics[width=.45\textwidth]{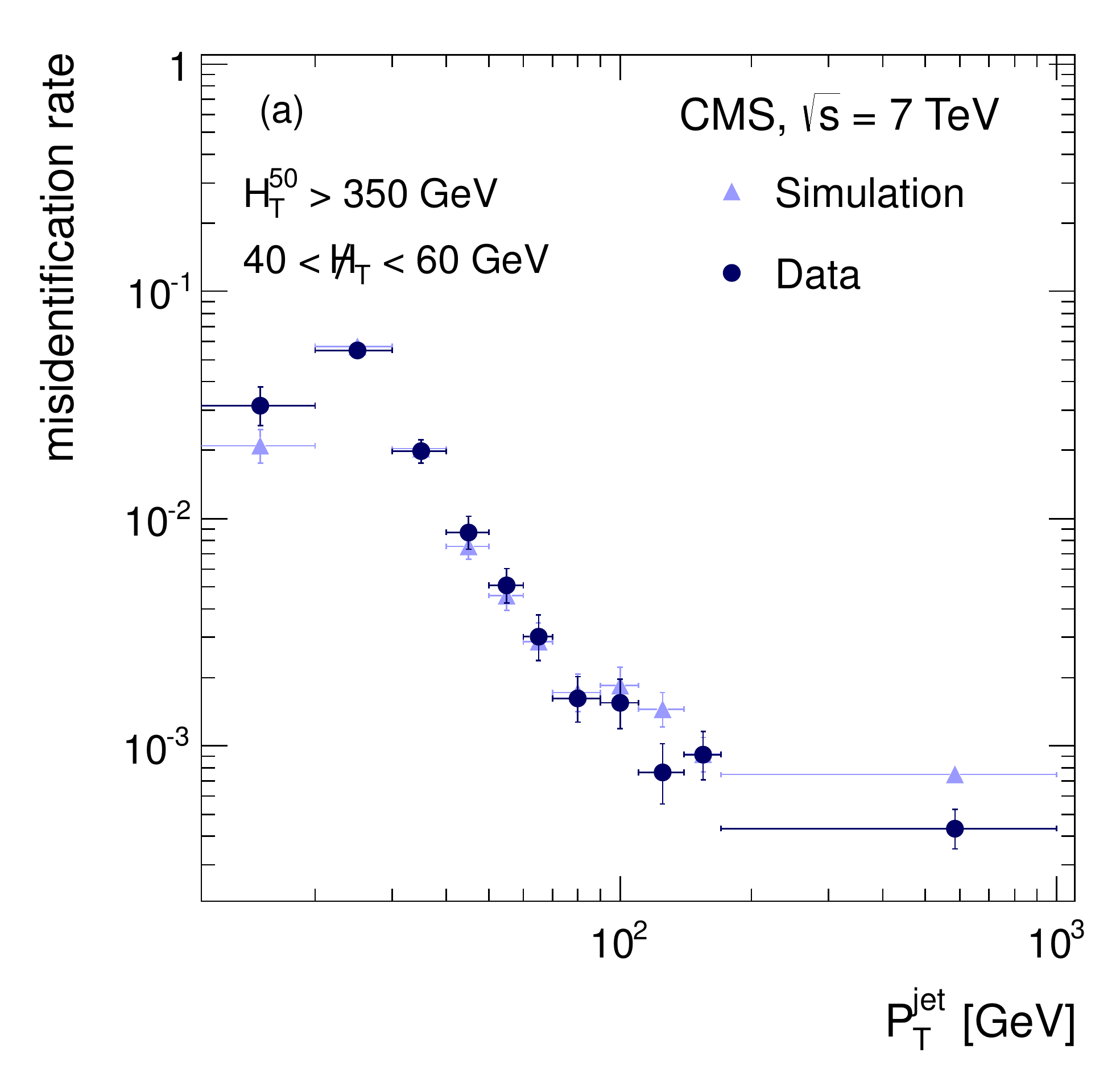}
\includegraphics[width=.45\textwidth]{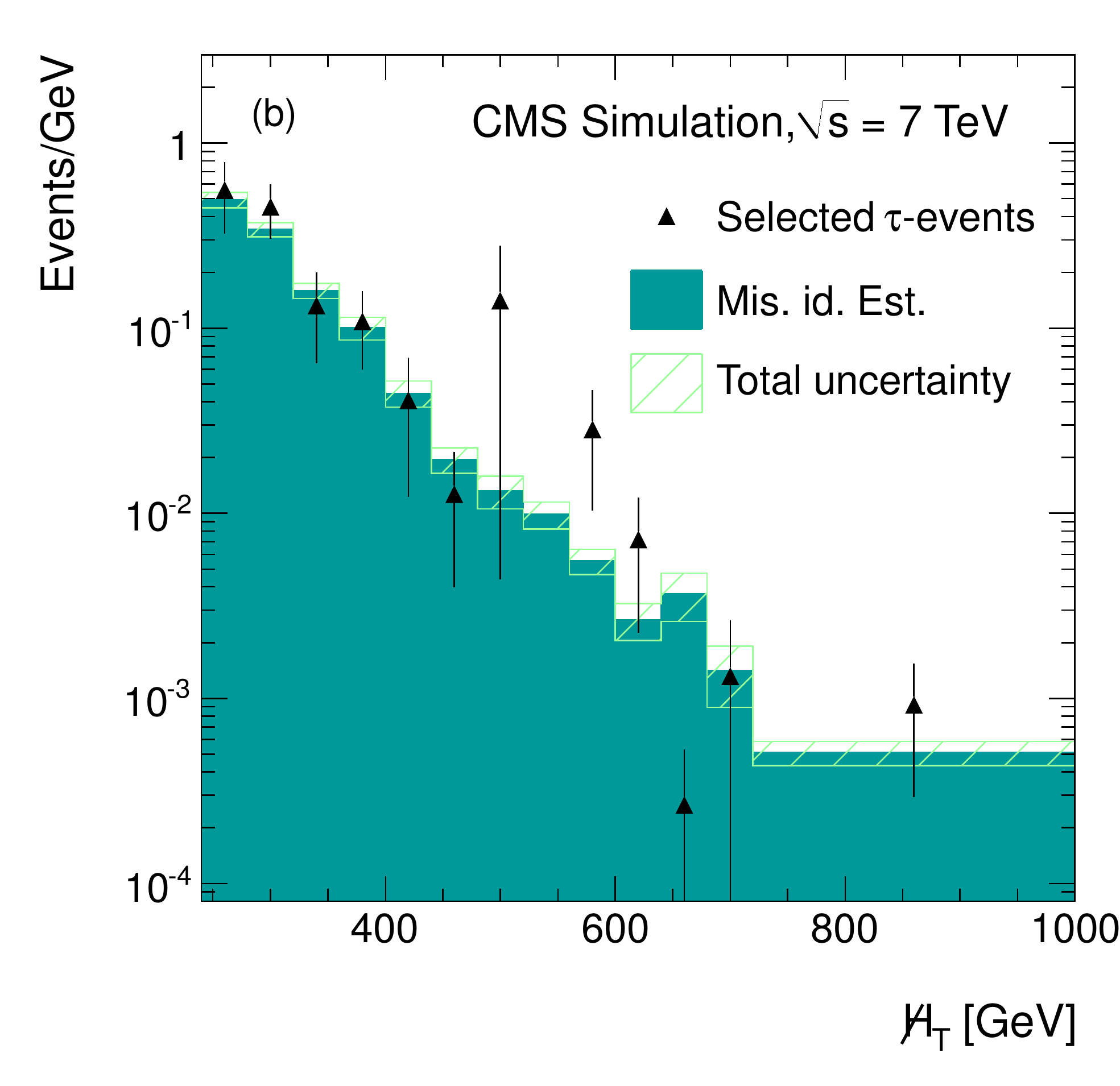}\\
\end{center}
\caption{(a) The rate of jet misidentification as a $\tauh$ lepton in simulation (triangular symbols) and data (circular symbols) as a
function of $\pt^\text{jet}$
for events with $\HT^{50}>350$\GeV and $40 < {\MHT} < 60$\GeV; (b) The \MHT\ distribution estimated in simulated events with
$\HT^{50}>350$\GeV, where the triangular symbols represent events that pass the baseline selection, the filled blue area shows the predicted
events, and the hatched area shows the total uncertainty on the prediction. These distributions correspond to the single-\tauh\ final state.}
\label{fig::fakeRatesComp}
\end{figure}
\begin{equation}
 w^{\text{corr}}_{\text{event}} = 1 - \prod_i^n (1-f_i),
\label{eq:singlefake}
\end{equation}
where $n$ is the number of jets.
The measured misidentification rates shown in Fig.~\ref{fig::fakeRatesComp}(a) are applied to data events in the region
with $\HT^{50}>350$\GeV and with $\HT^{50}>600$ for two regions of \MHT:
$60 < \MHT\ < 80$\GeV and $80 < \MHT\ < 100$\GeV. These four regions are dominated by QCD multijet events.
The results for data and simulation, as well as the predicted fraction of QCD multijet events, are
shown in Table~\ref{tab::QCDClosure}.
The ratio of selected events over predicted events is statistically
compatible with one and stable over the range of ${\MHT}$.
Figure~\ref{fig::fakeRatesComp}(b) shows the \MHT\ distributions of predicted and selected events for
simulated QCD multijet events with $\HT^{50}>350$\GeV.
The two distributions agree over the whole range of \MHT\ .

\begin{table*}[tbh]
\begin{center}
\topcaption{The percentage of QCD multijet events in the $\MHT$ binned samples for different QCD multijet dominated regions in the single-\tauh final
state.}
\label{tab::QCDClosure}
\begin{tabular}{cccc}
\hline
&\multicolumn{3}{c}{\MHT\ [\GeVns{}]}\\
$\HT^{50}>350$\GeV&60--80&80--100&$>250$\\
\hline
QCD fraction&97\%&93\%&6\%\\
selected/predicted (sim)&$0.98\pm0.06$&$0.96\pm0.07$&$1.24\pm0.28$\\
selected/predicted (data)&$1.01\pm0.08$&$0.88\pm0.13$&--\\ \hline
&\multicolumn{3}{c}{\MHT\ [\GeVns{}]}\\
$\HT^{50}>600$\GeV&&&$>$400\\  \hline
QCD fraction&96\%&93\%&17\%\\
selected/predicted (sim)&$0.94\pm0.09$&$0.85\pm0.09$&$2.43\pm1.45$\\
selected/predicted (data)&$1.14\pm0.26$&$0.97\pm0.37$&--\\ \hline
\end{tabular}
\end{center}
\end{table*}

\subsection{Estimate of backgrounds in the multiple-\texorpdfstring{$\tauh$}{tau(h)} final state}\label{sec:dataSets}

The estimate of the SM background contributions to the SR sample for multiple-$\tauh$ events is based on the number of observed events in CRs.
The events in each CR are selected with similar selection requirements to those used in
the SR, but
are enriched with events
from the background process in question.
Correction factors and selection efficiencies are measured in those CRs and used to extrapolate to the SR.
We use the observed jet multiplicity in each CR along with the measured rate at which a jet is misidentified
as a $\tauh$ to calculate the yield in the SR.
The following equation is used to estimate each background contribution B:

\begin{equation}
N_{\mathrm{B}}^\mathrm{SR} = N_{\mathrm{B}}^\mathrm{CR}[\alpha_{\tau\tau}{\mathcal{P}} (0) + \alpha_{\tau j}{\mathcal{P}} (1) + \alpha_{jj}{\mathcal{P}} (2)],
\label{eq:background}
\end{equation}

where $N_{\mathrm{B}}^\mathrm{SR}$ is the predicted rate in the SR,
$N_{\mathrm{B}}^\mathrm{CR}$ is the observed number of events in the CR, and
$\alpha_{xy}$ is the correction factor for acceptance and efficiency for events in the CR with true physics objects
``$x$" and ``$y$". Here the physics object can be a $\tauh$ or a quark or gluon jet.
Since the dominant SM backgrounds contribute to the SR when
jets are misidentified as $\tauh$ lepton, the background estimation strategy outlined in
Eq.~(\ref{eq:background}) relies on
the determination of the event probability ${\mathcal{P}} (m)$ for at least ``$m$" jets to be misidentified as a
$\tauh$, where ${\mathcal{P}} (m)$ is the product of three factors:
(i) the probability $P(N)$ for an event to contain $N$ jets,
(ii) the number of possible ways for exactly $n$ jets to pass the $\tauh$ identification criteria given
$N$ possible jets $C(N,n)=N!/n!(N-n)!$, and (iii) the probability $f$ for a single jet to be misidentified as a
$\tauh$. The ${\mathcal{P}} (m)$ terms are given by:
\begin{equation}
{\mathcal{P}} (m)=\sum_{N=m}^{\infty}P(N)\sum_{n=m}^{N}C(N,n)f^{n}(1-f)^{N-n}.
\label{eq:EventProb}
\end{equation}

Equation~(\ref{eq:EventProb}) would be identical to Eq.~(\ref{eq:singlefake})
if used in the case of the single-\tauh\ final state.
Equation~(\ref{eq:background}) is used to estimate the \ttbar, $\PW+\text{jets}$, and $\cPZ+\text{jets}$ background contributions to the
SR. The $P(N)$ terms are determined from data using the jet multiplicity distribution in each CR, while the $f$ terms
are measured for each background process by determining
the fraction of jets in each CR that pass the $\tauh$ identification criteria.
Since the QCD multijet contribution to the SR for the multiple-$\tauh$ final state is negligible according to simulation,  a data-to-MC scale factor
is used to correct the QCD multijet prediction from simulation. In the sections that follow, the selections used to define high purity CRs are outlined
and the correction factors $\alpha_{xy}$ used in Eq.~(\ref{eq:background}) are defined.
The fraction of events with two $\tauh$ leptons is denoted $A_{\tau \tau}$, the fraction with one $\tauh$ lepton and one jet misidentified as a
$\tauh$ lepton is denoted $A_{\tau j}$, and the fraction with two jets
misidentified as $\tauh$ leptons is denoted $A_{jj}$.

\subsubsection{Estimate of the \texorpdfstring{\ttbar}{t t-bar} event background to the multiple-\texorpdfstring{$\tauh$}{tau(h)} final state}

\begin{figure*}[htb]
\begin{center}
\includegraphics[width=.45\textwidth]{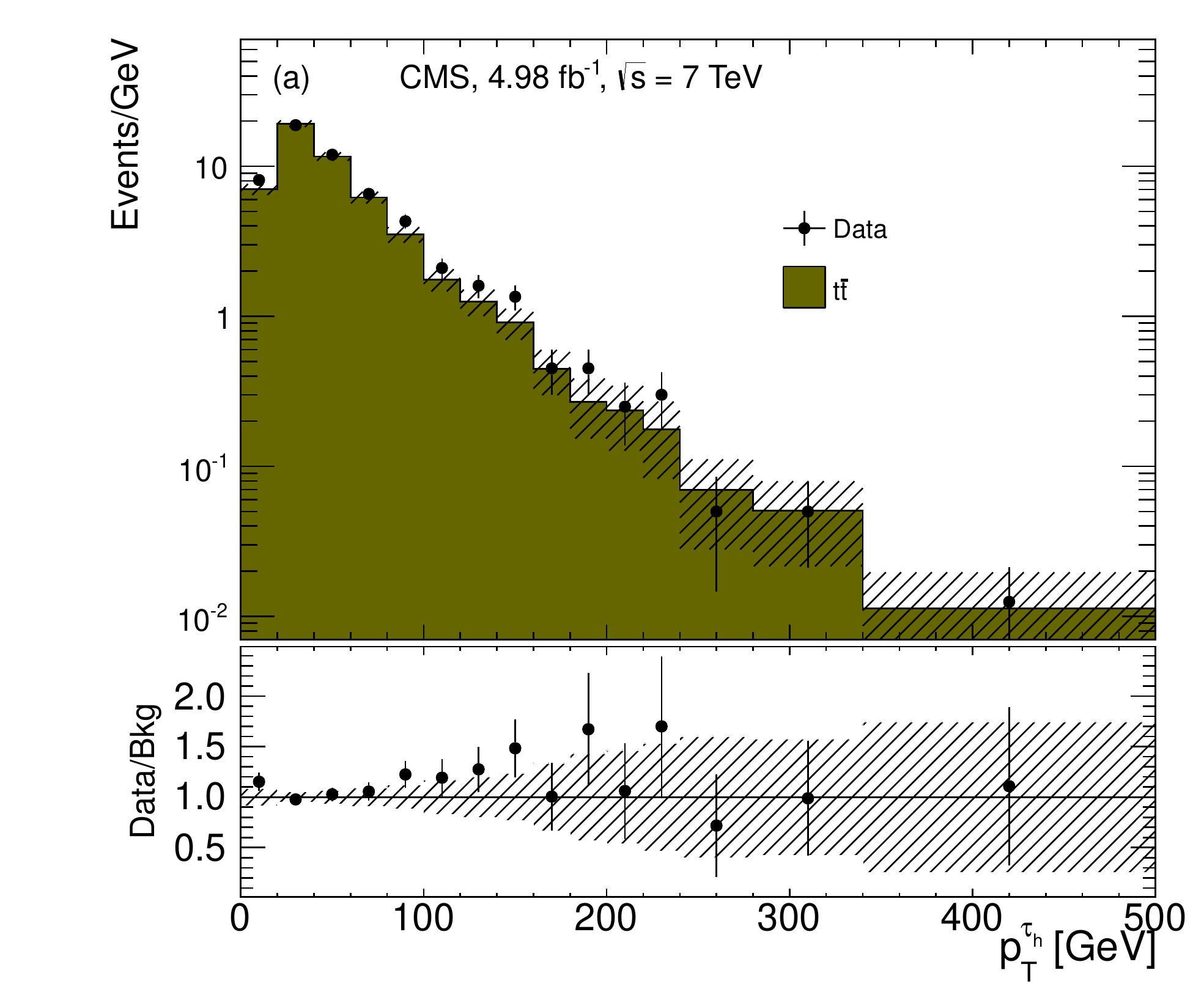}
\includegraphics[width=.45\textwidth]{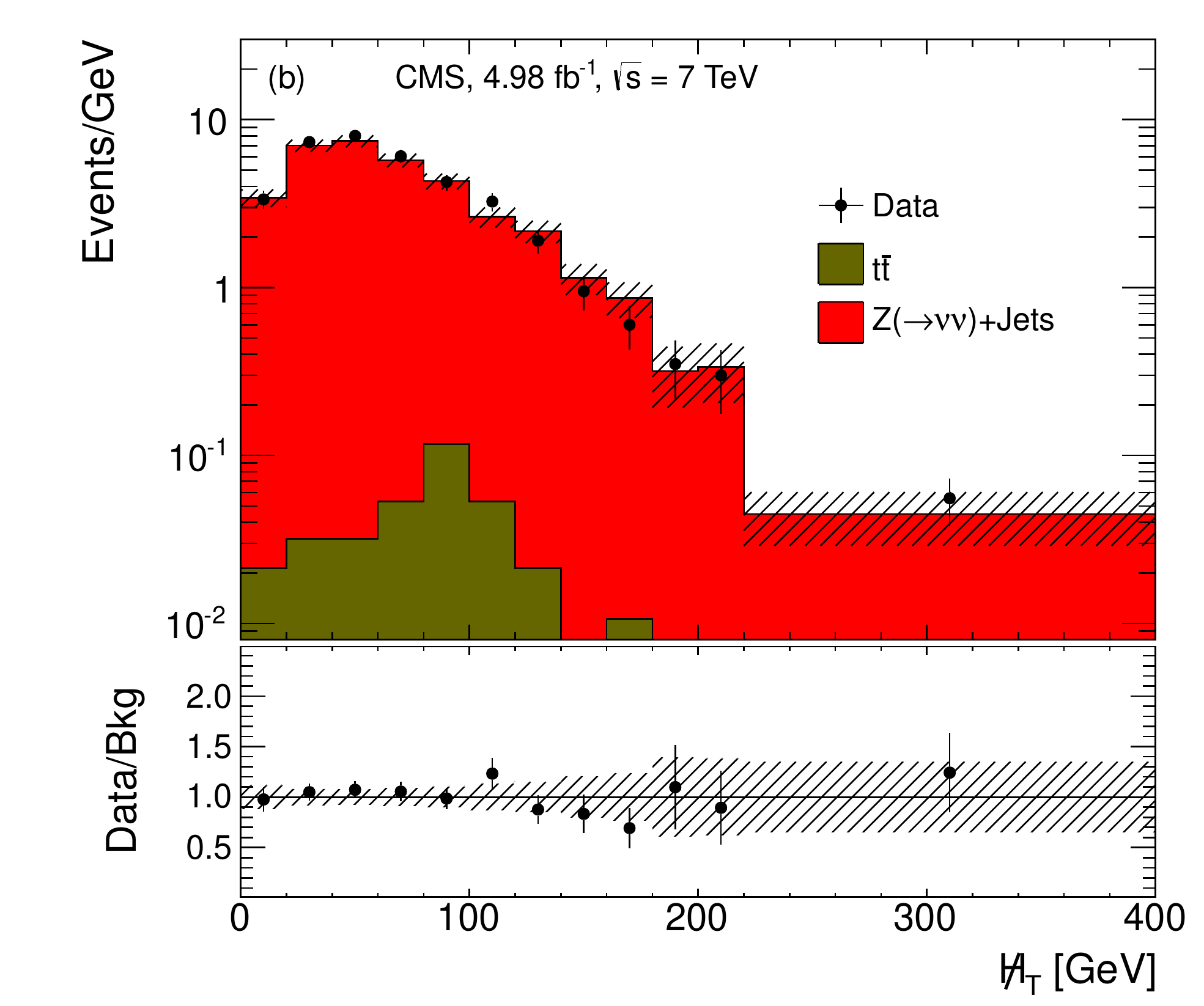}
\includegraphics[width=.45\textwidth]{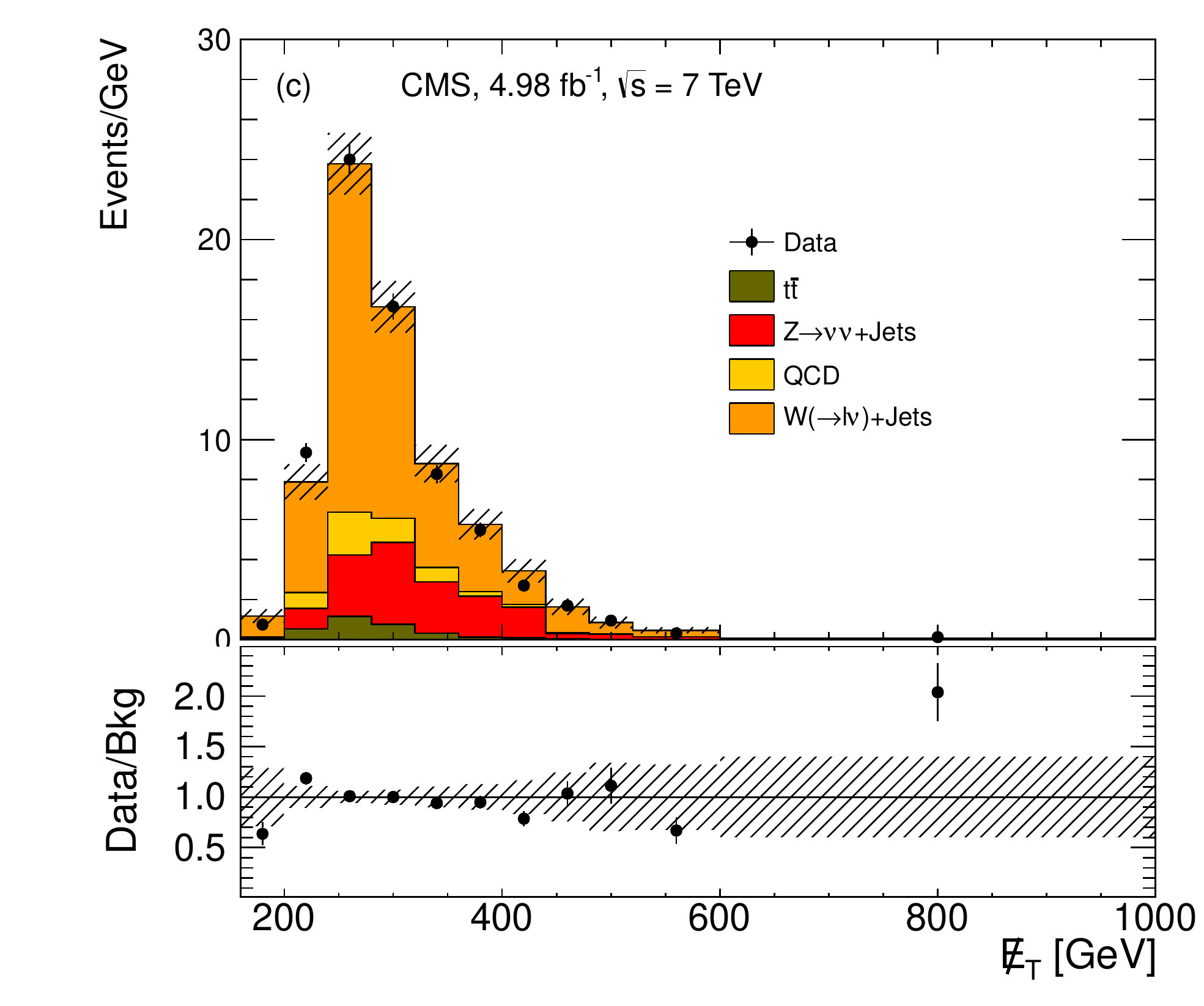}
\includegraphics[width=.45\textwidth]{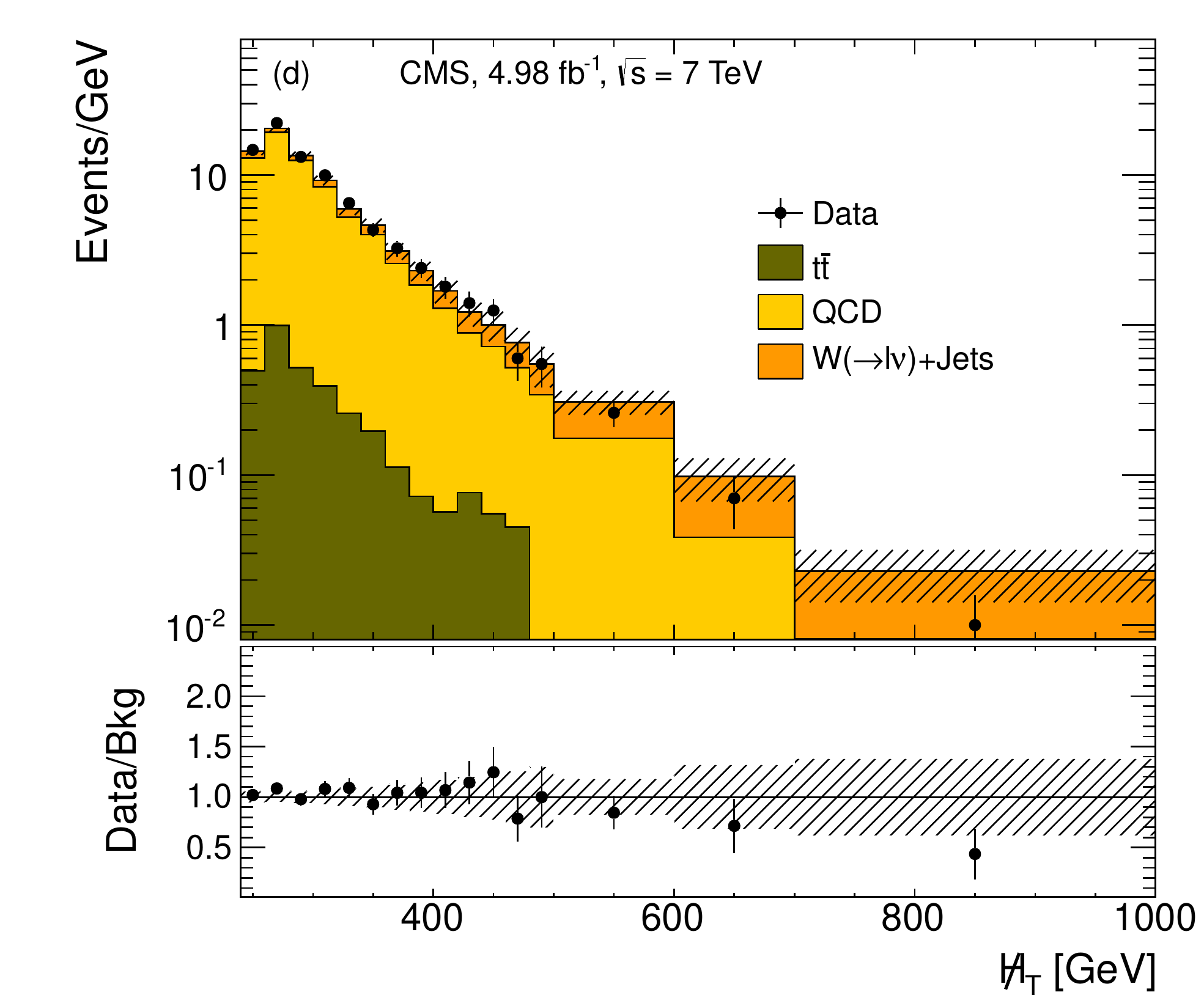}
\end{center}
  \caption{Data-to-MC comparison for the multiple-\tauh\ final state: (a) the \pt\ distribution of the \tauh\ candidate in the \ttbar CR;
(b) ${\MHT}$ distribution in the $\cPZ (\to \mu\mu)$ + jets CR;
(c) ${\ETslash}$ distribution in the $\PW+\text{jets}$ CR; and
(d) ${\MHT}$ distribution with the requirement $|\Delta \phi(j_{2},{\MHT})|<0.1$.
The bottom panes show the ratio between data and background while the hatched area depicts the total uncertainty on the MC.}
  \label{fig::SignalMassPlot1}
\end{figure*}

To estimate the contribution of \ttbar events to the multiple-\tauh\ SR, a CR is selected
by removing the $\tauh$ isolation requirement and by
requiring the presence of at least two b-quark jets (b jets), identified using the track-counting-high-efficiency (TCHE) algorithm at the medium working point
\cite{BTV-11-001}.
Because QCD multijet, $\PW+\text{jets}$, ${\cPZ} (\to \tau\tau)$ + jets and ${\cPZ} (\to \nu\nu)$ + jets
events are unlikely to contain two $\cPqb$ jets, this requirement provides a sample
in which about 99\% of the events are \ttbar events, according to simulation.
Figure~\ref{fig::SignalMassPlot1}(a) shows the \pt\ distribution of $\tauh$ leptons in the \ttbar CR for data and simulation.

According to simulation, the fraction of events in the \ttbar control sample that contains one
genuine $\tauh$ is $A_{\tau j}=0.166 \pm 0.011$, while the fraction without a genuine $\tauh$ is $A_{jj}=0.834 \pm 0.025$.
The genuine $\tauh\tauh$ contribution is negligible ($A_{\tau \tau}\sim 0$) according to simulation.
Incomplete knowledge of the genuine $\tauh\tauh$ contribution is included as a source of systematic uncertainty in the
\ttbar background prediction. Therefore, $\alpha_{\tau j}$ in Eq.~(\ref{eq:background}) is given by
$A_{\tau j}{\varepsilon_{\tau}^{\text{iso}}}/{P(\text{2 \cPqb\ jets})}$, where $\varepsilon_{\tau}^{\text{iso}}$ is
the probability for a $\tauh$ lepton to pass the isolation requirement, while $\alpha_{jj}$ is given by
${A_{jj}}/{P(2\,\cPqb\,\text{jets})}$. The probability $P(\text{2 \cPqb\ jets})$ to identify
two or more $\cPqb$ jets is determined by the $\cPqb$ jet identification efficiency factor
\cite{BTV-11-001}. The number of \ttbar events in the SR is calculated as:
\begin{equation}
N_{\ttbar}^\mathrm{SR} =
\frac{N^\mathrm{CR}_{t \overline{t}}}{P(\text{2 \cPqb jets})}[ A_{\tau j}\varepsilon_{\tau}^{\text{iso}}{\mathcal{P}} (1) + A_{j j}{\mathcal{P}} (2) ].
\label{eq:ttbarFakeRate}
\end{equation}
The probability for a jet in a \ttbar event to be misidentified as a $\tauh$ lepton has an average measured value of
$f = 0.022 \pm 0.004$.
Cross checks are made to validate the use of the $\cPqb$-jet identification efficiency as measured in Ref.~\cite{BTV-11-001} for this analysis.
The estimated \ttbar contribution in the SR is determined to be $N_{\ttbar}^\mathrm{SR} = 2.03 \pm 0.36$.

\subsubsection{Estimate of the ${\cPZ} (\to \nu\nu)$ + jets event background to the multiple-$\tauh$ final state}

The contribution of $\cPZ (\to \nu\nu)$ + jets events to the multiple-$\tauh$ SR is evaluated by selecting a sample of
$\cPZ (\to \mu \mu)$ + jets events and treating the muons as neutrinos.
The sample is collected using a trigger designed to select a muon and a $\tauh$.
Jet selection criteria similar to those used for the SR sample are imposed.
In addition, we require two muons passing the criteria outlined in Section~\ref{sec:leptonRecoId}.
The control sample
has a purity of about $99$\% as estimated from simulation.
The ${\MHT}$ distribution for events in this CR is shown in Fig.~\ref{fig::SignalMassPlot1}(b).
The $\cPZ (\to \nu \nu)$ + jets background is estimated by interpreting the \pt\
of the pair of muons as ${\MHT}$. In order to predict the $\cPZ (\to \nu \nu)$ + jets rate in the
SR, the $\cPZ (\to \mu \mu)$ + jets sample is corrected for the ratio of the branching fractions
$R=B(\cPZ\to \nu\nu)/B(\cPZ\to\mu\mu)$, for trigger efficiencies, for the geometric acceptance
$A_{\mu}$ as measured from simulation, and for the reconstruction efficiency
$\varepsilon_{\mu}^{\text{reco}}$ as measured from data.
Therefore, $\alpha_{jj}$ in Eq.~(\ref{eq:background}) is given by:
\begin{equation}
\frac{1}{A_{\mu}^{2}\varepsilon_{\mu}^{\text{reco 2}}}\frac{B(\cPZ \to \nu \nu)}{B(\cPZ
\to \mu \mu)}\frac{\varepsilon_{{\MHT}}^{\text{Trigger}}}{\varepsilon_{\mu\tau}^{\text{Trigger}}}
\varepsilon^{{\MHT}}.
\end{equation}

Since there is no prompt production of a genuine $\tauh$ in the $\cPZ (\to \mu \mu)$ + jets sample, $\alpha_{\tau j}=0$
and $\alpha_{\tau \tau}=0$.
The $\cPZ (\to \nu\nu)$ + jets contribution to the SR is calculated as:
\begin{equation}
N_{\cPZ \to \nu \nu + \text{jets}}^\mathrm{SR} =
\frac{N^\mathrm{CR}_{\cPZ \to \mu \mu + \textrm{jets}}}{A_{\mu}^{2}\varepsilon_{\mu}^{\text{reco 2}}}
R
\frac{\varepsilon_{{\MHT}}^{\text{Trigger}}}{\varepsilon_{\mu\tau}^{\text{Trigger}}}
\varepsilon^{{\MHT}}
{\cal{P}} (2),
\label{eq:ZnunuFakes}
\end{equation}
where $\varepsilon^{\text{Trigger}}_{{\MHT}}$ is the ${\MHT}$ trigger efficiency
and $\varepsilon^{\text{Trigger}}_{\mu\tau}$ the $\mu\tauh$ trigger efficiency. The efficiency for
the ${\MHT}>250$\GeV signal selection ($\varepsilon^{{\MHT}}$)
is determined by calculating the fraction of
the observed events in the CR that have ${\MHT}>250$\GeV.
The muon identification efficiency $\varepsilon_{\mu}$ is measured using a
``tag-and-probe" method.
The probability for a jet to be misidentified as a $\tauh$ lepton has a measured value of $f=0.016 \pm 0.002$.
The estimated $\cPZ (\to\nu\nu)$ + jets contribution to the SR is
determined to be $N_{\cPZ (\to\nu\nu)}^\mathrm{SR} = 0.03 \pm 0.02$.

\subsubsection{Estimate of the $\cPZ (\to \tau\tau)$ + jets event background to the multiple-$\tauh$ final state}

The contribution from $\cPZ \to \tau\tau$ events is determined with the
$\cPZ (\to \mu \mu)$ + jets CR sample used to estimate the background from $\cPZ (\to \nu \nu)$ + jets,
with the muons treated as $\tauh$ leptons.
The $\alpha_{xy}$ factors are more difficult to estimate for $\cPZ \to \tau\tau$ events since there are several ways
in which $\cPZ \to \tau\tau$ events can contribute to the SR: (i) both $\tauh$ leptons pass the kinematic acceptance and
identification criteria; (ii) both $\tauh$ leptons pass the kinematic acceptance criteria, but only one passes the identification
criteria; (iii) one $\tauh$ fails the kinematic acceptance criteria, while the other $\tauh$ passes both the kinematic
acceptance and identification criteria; or (iv) both $\tauh$ leptons fail the kinematic acceptance criteria.
The $\cPZ (\to \tau \tau) + \text{jets}$ contribution to the SR is calculated as:
\begin{equation}\begin{split}
N_{\cPZ \to \tau \tau}^\mathrm{SR} =
&N^\mathrm{CR}_{\cPZ \to \mu \mu}R \Bigg[
\frac{A_{\tau}^{2}\varepsilon_{\tau}^{2}}{A_{\mu}^{2}\varepsilon_{\mu}^{\text{reco 2}}} +
\frac{2A_{\tau}^{2}\varepsilon_{\tau}(1-\varepsilon_{\tau})}{A_{\mu}^{2}\varepsilon_{\mu}^{\text{reco 2}}}{\cal{P}} (1) +\\
&\frac{2A_{\tau}(1-A_{\tau})\varepsilon_{\tau}}{A_{\mu}^{2}\varepsilon_{\mu}^{\text{reco 2}}}{\mathcal{P}} (1) +
\frac{(1-A_{\tau})^{2}}{A_{\mu}^{2}\varepsilon_{\mu}^{\text{reco 2}}}{\mathcal{P}} (2) \Bigg],
\label{eq:ZtautauFakes}
\end{split}\end{equation}
where $R$ is given by:
\begin{equation}
\frac{B(\cPZ \to \tau \tau)B^{2}(\tau \to \tauh)}{B(\cPZ \to \mu \mu)}
\frac{\varepsilon_{{\MHT}}^{\text{Trig}}}{\varepsilon_{\mu\tau}^{\text{Trig}}}\varepsilon^{{\MHT}},
\end{equation}
$A_{\tau}$ is the $\tauh$ acceptance,
$\varepsilon_{\tau}$ is the $\tauh$ identification efficiency in this control sample, and $f=0.016 \pm 0.002$.
The estimated Z ($\to\tau\tau$) + jets contribution to the SR is
determined to be $N_{{\cPZ} (\to\tau\tau)}^\mathrm{SR} = 0.21 \pm 0.13$.

\subsubsection{Estimate of the W + jets event background to the multiple-$\tauh$ final state}

To select the $\PW+\text{jets}$ CR, the $\tauh$ isolation requirement, which discriminates between a $\tauh$ lepton and other jets, is removed from the SR
selection requirements.
However, the lack of the $\tauh$ isolation requirement increases the contribution from other backgrounds as most of the backgrounds arise
because jets are misidentified as a $\tauh$ lepton.
To minimize the contribution from \ttbar production, events are required to have no jets identified as a $\cPqb$ jet.
This requirement reduces the contamination from \ttbar events to around $5$\%.
The purity of the $\PW+\text{jets}$ CR is approximately $65$\%.
Figure~\ref{fig::SignalMassPlot1}(c) shows the ${\ETslash}$ distribution, defined as the magnitude of the negative of the vector sum of the transverse
momentum of all PF objects in the event, for events in the $\PW+\text{jets}$ CR.
The contributions of QCD multijet, \ttbar, and $\cPZ (\to \nu \nu)+\text{jets}$
events are subtracted in order to determine the number of $\PW+\text{jets}$ events
in the CR. The predicted rates for QCD multijet,
\ttbar, and $\cPZ (\to \nu \nu)$ + jets events are determined by
extrapolating from their corresponding CRs.
Since there is no genuine multiple-$\tauh$ production in $\PW+\text{jets}$, $\alpha_{\tau \tau}=0$.
According to simulation, the fraction of events in the CR with one
genuine $\tauh$ is $A_{\tau j}=0.149 \pm 0.016$, while the fraction of events without a genuine $\tauh$ is $A_{jj}=0.851 \pm 0.038$.
Therefore, $\alpha_{\tau j}$ in Eq.~(\ref{eq:background}) is given by
$A_{\tau j}{\varepsilon_{\tau}^{\text{iso}}}/{P(\text{0 \cPqb\ jets})}$, where $\varepsilon_{\tau}^{\text{iso}}$ is
the probability for a $\tauh$ to pass the isolation requirement and $P(\text{0 \cPqb\ jets})$ is the probability
to not have any light-quark or gluon jet misidentified as a b jet. Similarly, $\alpha_{jj}$ is given by
${A_{jj}}/{P(\text{0 \cPqb\ jets})}$.
The contribution of $\PW+\text{jets}$ events to the SR is then calculated as:

\begin{equation}
N_{W+\text{jets}}^\mathrm{SR}      =
\frac{N^{\text{After subtraction}}_{\PW+\textrm{jets}}}{P(\text{0 \cPqb\ jets})}\left[  A_{\tau j}\varepsilon_{\tau}^{\text{iso}}{\mathcal{P}} (1) + A_{jj}{\mathcal{P}} (2) \right].
\label{eq:wjetsFakeRate}
\end{equation}

The average rate at which jets are misidentified as a $\tauh$ lepton is measured to be $0.019 \pm 0.001$.
The rate $f_\cPqb$ at which light-quark jets or gluon jets are misidentified as a b jet is used to
determine $P(\text{0 \cPqb\ jets})$.
The estimated $\PW+\text{jets}$ contribution to the SR is determined to be $N_{\PW + \textrm{jets}}^\mathrm{SR} = 5.20 \pm 0.63$.

\subsubsection{Estimate of the QCD multijet event background to the multiple-$\tauh$ final state}

QCD multijet events contribute to the multiple-$\tauh$ SR when mismeasurements of jet energies lead to large values of
${\MHT}$ and when jets are misidentified as $\tauh$ candidates. By removing the $\tauh$
isolation criteria and inverting the $|\Delta \phi(j_{2},{\MHT})|$ requirement,
a QCD CR sample with about 99\% purity is obtained.
Figure~\ref{fig::SignalMassPlot1}(d) shows the expected and observed ${\MHT}$ distributions for this sample.
A scale factor is obtained from this CR and used to correct the signal prediction for QCD multijet events in simulation. The estimated contribution to the SR from
QCD multijet events is determined to be $N_\mathrm{QCD}^\mathrm{SR} = 0.02 \pm 0.02$.

\section{Systematic uncertainties}
\label{sec:systematics}

Systematic uncertainties are taken into account for both signal and background events and
are described separately. Both the signal and background are affected by the systematic
uncertainty in the identification of the \tauh\ candidate.
The systematic uncertainty for $\tauh$ identification is obtained using a
$\cPZ \to \tau\tau$ enhanced region and by correcting this cross section by that measured
for $\cPZ \to {\rm ee}$ and $\cPZ\to\mu\mu$ events. This uncertainty is validated on a control sample of $\cPZ \to \tau\tau$
events.
The level of agreement between data and simulation is found to be at the level of 7\%. Further validation of the performance of
$\tauh$ identification in a SUSY-like environment is performed by selecting
a \PW($\to\tau\nu\to\tauh\nu\nu$) + jets CR with large hadronic activity ($\HT$)
and large transverse momentum imbalance (${\MHT}$). The level of agreement between
the predicted rate for $\PW(\to\tau\nu\to\tauh\nu\nu)$ events and the observed number of events is within
7\% and is determined as a function of $\HT$ and ${\MHT}$.

\subsection{Systematic uncertainties on background events}
The principal sources of systematic uncertainty on the background predictions arise from the correction factors,
the finite number of events in the CRs, the measured rates at which jets are misidentified as a $\tauh$ lepton, and the
level of agreement between the observed and predicted numbers of events in CRs.

The contributions to the uncertainties on the correction factors are different for each background
category. The dominant effect is due to the uncertainty in the $\tauh$
identification efficiency.
In the multiple-$\tauh$ final state, uncertainties in the jet-energy scale (JES) \cite{ref:jes} and the $\tauh$-energy
scale (TES) \cite{tes} are used to evaluate how changes in
$\HT$, ${\MHT}$, and jet kinematics affect the
correction factors. The systematic uncertainty on the correction factors
due to the JES and TES is at most $\sim$3\% for all backgrounds.
Smaller contributions to the uncertainties in the
correction factors arise from the muon reconstruction and isolation efficiency ($<$1\%), the uncertainty
in the branching fractions ($\ll$1\%), and the uncertainties in trigger efficiency (1\%).

The systematic uncertainties on the
measured rates for jet misidentification as a $\tauh$ lepton
are dominated by the size of the jet sample
used to measure these rates and range from  2\% for the single-$\tauh$ final state to
5.6--10\% for the multiple-$\tauh$ final state. The level of agreement between the
observed and predicted number of events in MC studies of the CRs is used to assign an additional systematic
uncertainty and ranges from 2\%
for the single-$\tauh$ final state to 3\% for the multiple-$\tauh$ final state. Finally, the
systematic uncertainty arising from statistical uncertainties on the number of events in the
CRs ranges from 2--5\% for the multiple-$\tauh$ final state to 3--10\% for the
single-$\tauh$ final state.

\subsection{Systematic uncertainties on signal events}\label{sec:sigsystematics}

The main sources of systematic uncertainties in the SR are due to trigger efficiencies,
identification efficiencies, the energy, and momentum scales, the luminosity
measurement and PDFs.
The uncertainty on the luminosity measurement is 2.2\% \cite{ref:lumiNew}.
Systematic uncertainties on the ${\MHT}$ triggers (2.5\%) are measured
using a sample in which around 99\% of the events are \ttbar events, which have a similar topology to
events in the SR samples. The systematic uncertainties on the TES and JES (3.0\%)
yield an uncertainty on the signal acceptance of 2.3\%. The uncertainty on the
\MHT\ scale depends on the uncertainty of the JES
(2--5\% depending on the $\eta$ and \pt\ values of the jet) and on the unclustered energy scale
(10\%). Unclustered energy is defined as the energy found ``outside'' any reconstructed
lepton or jet with $\pt > 10$\GeV. The unclustered energy scale uncertainty has a negligible
systematic uncertainty on the signal acceptance.
The systematic uncertainty due to imprecise knowledge of the PDFs (11\%)
is determined by comparing the CTEQ6.6L\cite{Nadolsky:2008fk}, MSTW 2008 NLO \cite{MSTW2008NLO}, and
NNPDF2.1 \cite{NNPDF21} PDFs with the default PDF \cite{PDF4LHC}.
The systematic uncertainty due to the imprecise modeling of the initial-state and final-state
radiation \cite{partonShower} is negligible ($\ll$1\%).
The systematic uncertainties associated with event pileup are also negligible.
Uncertainties on the theoretical cross sections are evaluated by varying the PDFs and by changing the renormalization and factorization scales
by a factor of two \cite{ref:CMSSMuncerts, ref:CMSSMuncerts2, ref:CMSSMuncerts3, ref:CMSSMuncerts4, ref:CMSSMuncerts5, ref:CMSSMuncerts6}.

\section{Results}

\begin{figure}[htb]
\begin{center}
\includegraphics[width=.45\textwidth]{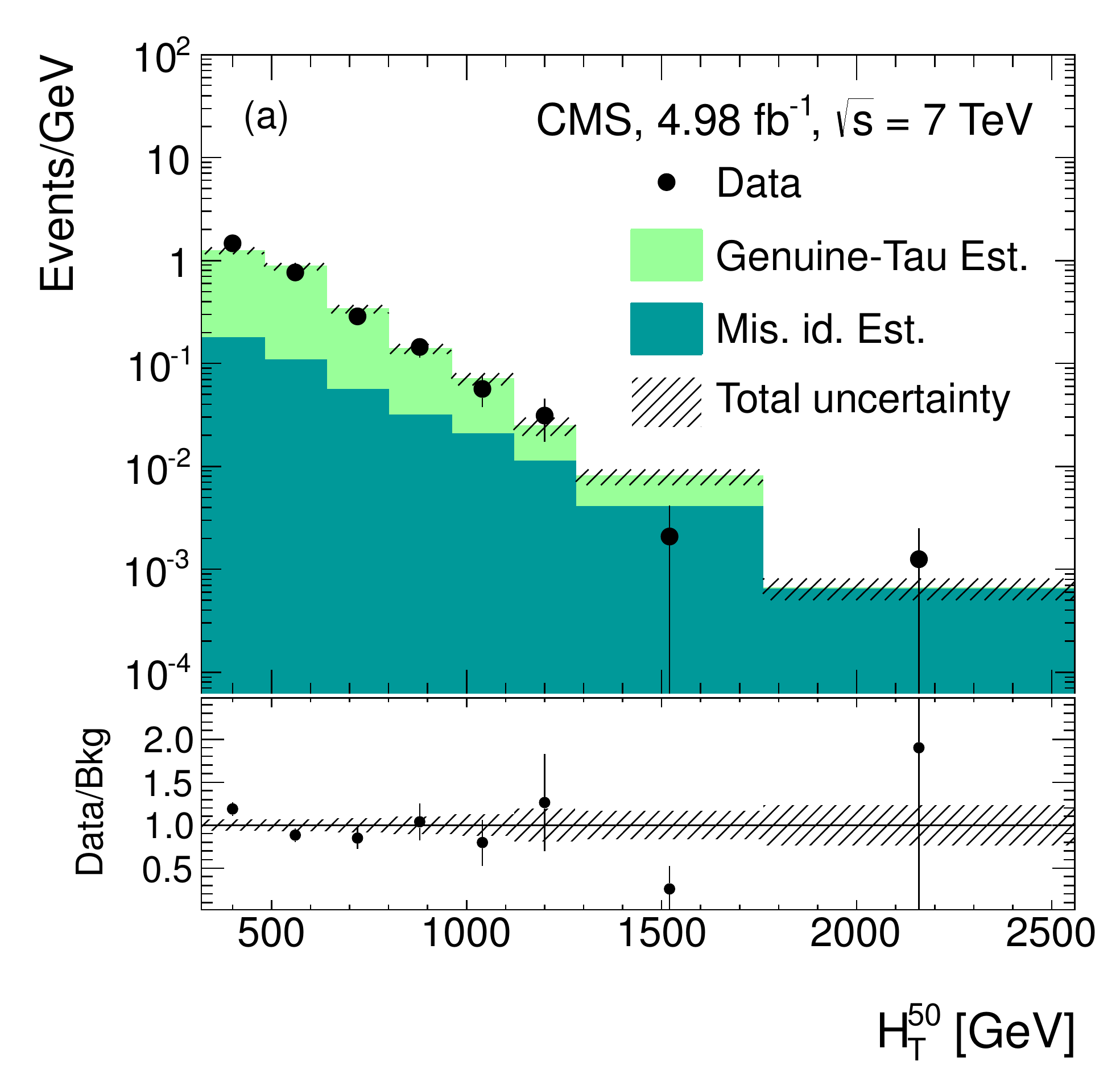}
\includegraphics[width=.45\textwidth]{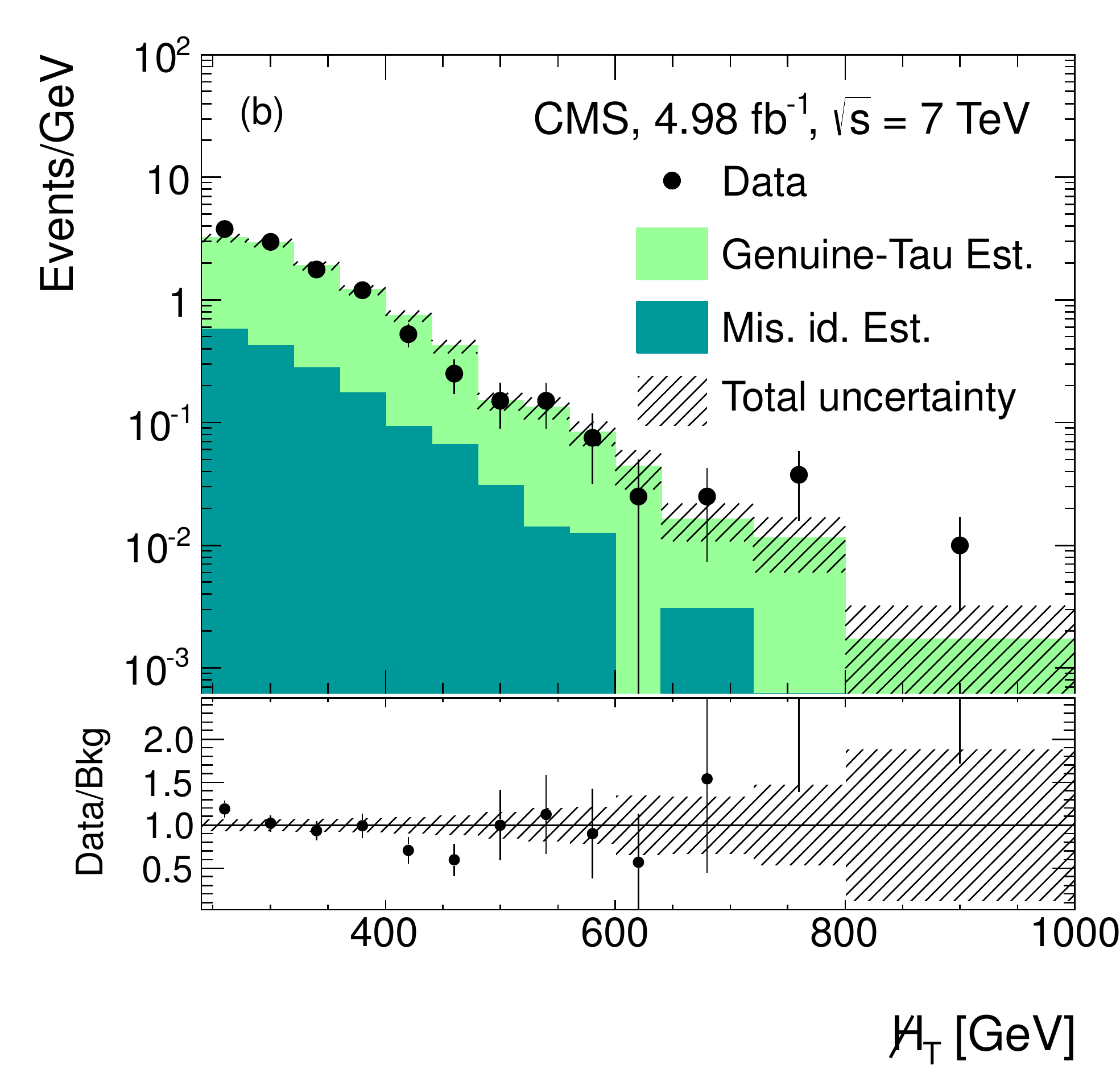}
\end{center}
 \caption{Distributions of (a) $\HT^{50}$, and (b) \MHT\ for the single-$\tauh$ final state. The points with errors represent data that satisfy
the baseline selection while the filled green (light) and filled blue (dark) areas shows the predicted backgrounds due to events containing a
genuine \tauh\ and a misidentified $\tauh$, respectively. The hatched area shows the total uncertainty on the prediction.}
 \label{fig::dataControl1}
\end{figure}

For the single-\tauh\ final state, the number of background events containing
a genuine $\tauh$, as well as the number of background events containing a misidentified
$\tauh$, are estimated with data.
The results for the baseline and the full selection are listed in Table~\ref{tab::results}.
Figure~\ref{fig::dataControl1} shows the $\HT^{50}$\ and \MHT\ distributions of data and the different background predictions.
The observed number of events in data is in agreement with
the SM predictions.

\begin{table}[tbh]
\begin{center}
\topcaption{Number of data and estimated background events with statistical and systematic uncertainties, respectively, in the single-$\tauh$ final state.} 
\begin{tabular}{ccc}
\hline
Process & Baseline & Signal Region \\ \hline
Fake$-\tauh$ & $67 \pm 2  \pm 19 $ & $3.4 \pm 0.4  \pm 1.0 $ \\
Real$-\tauh$ & $367 \pm 10  \pm 27 $ & $25.9 \pm 2.5  \pm 2.3 $ \\
Estimated $\sum SM$ & $434 \pm 10  \pm 33 $ & $ 29.3 \pm 2.6  \pm 2.5  $ \\ \hline
Data & $444$ & $28$\\ \hline
\end{tabular}
\label{tab::results}
\end{center}
\end{table}

\begin{figure}[htb]
\begin{center}
  \includegraphics[width=.45\textwidth]{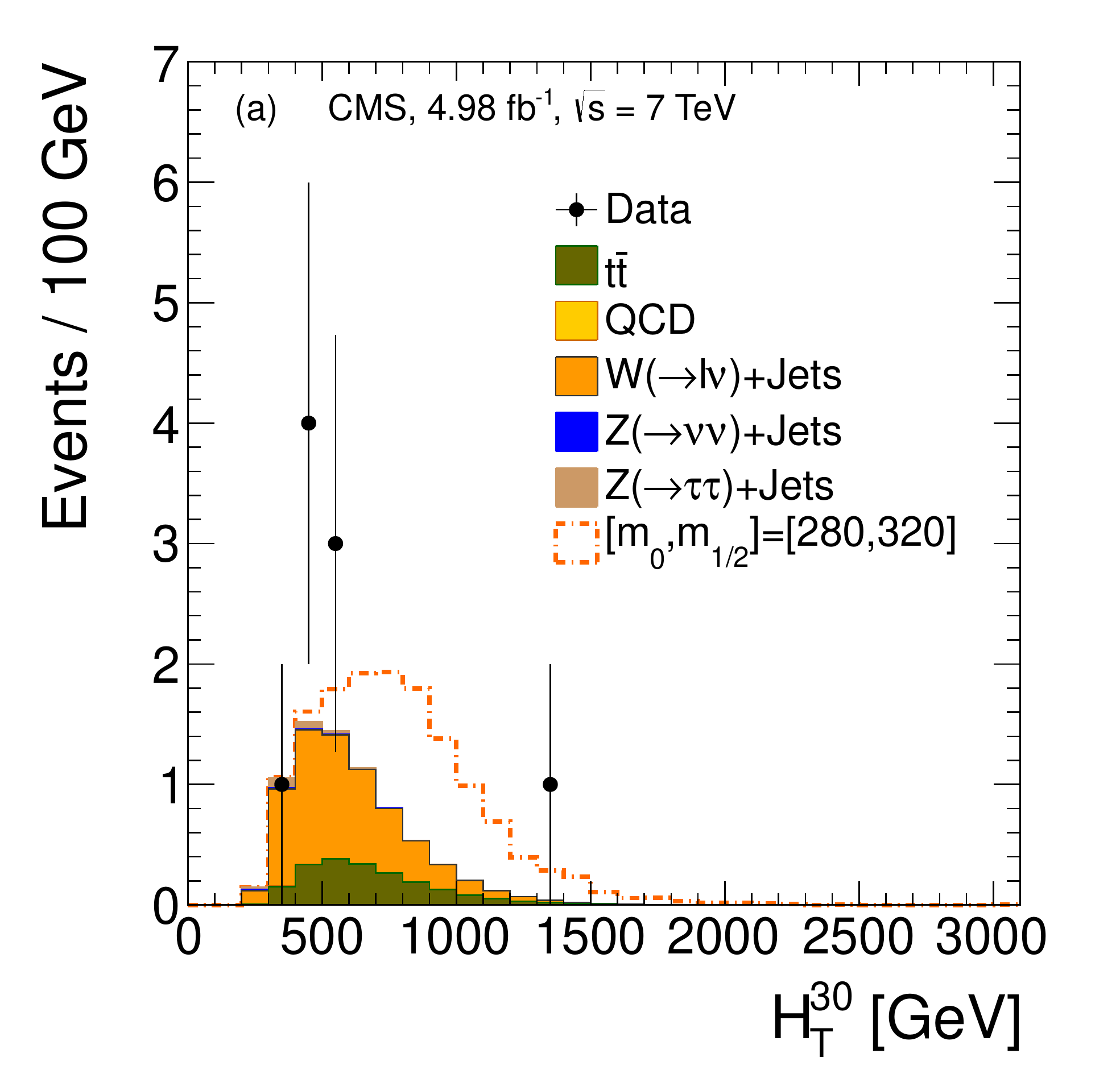}
  \includegraphics[width=.45\textwidth]{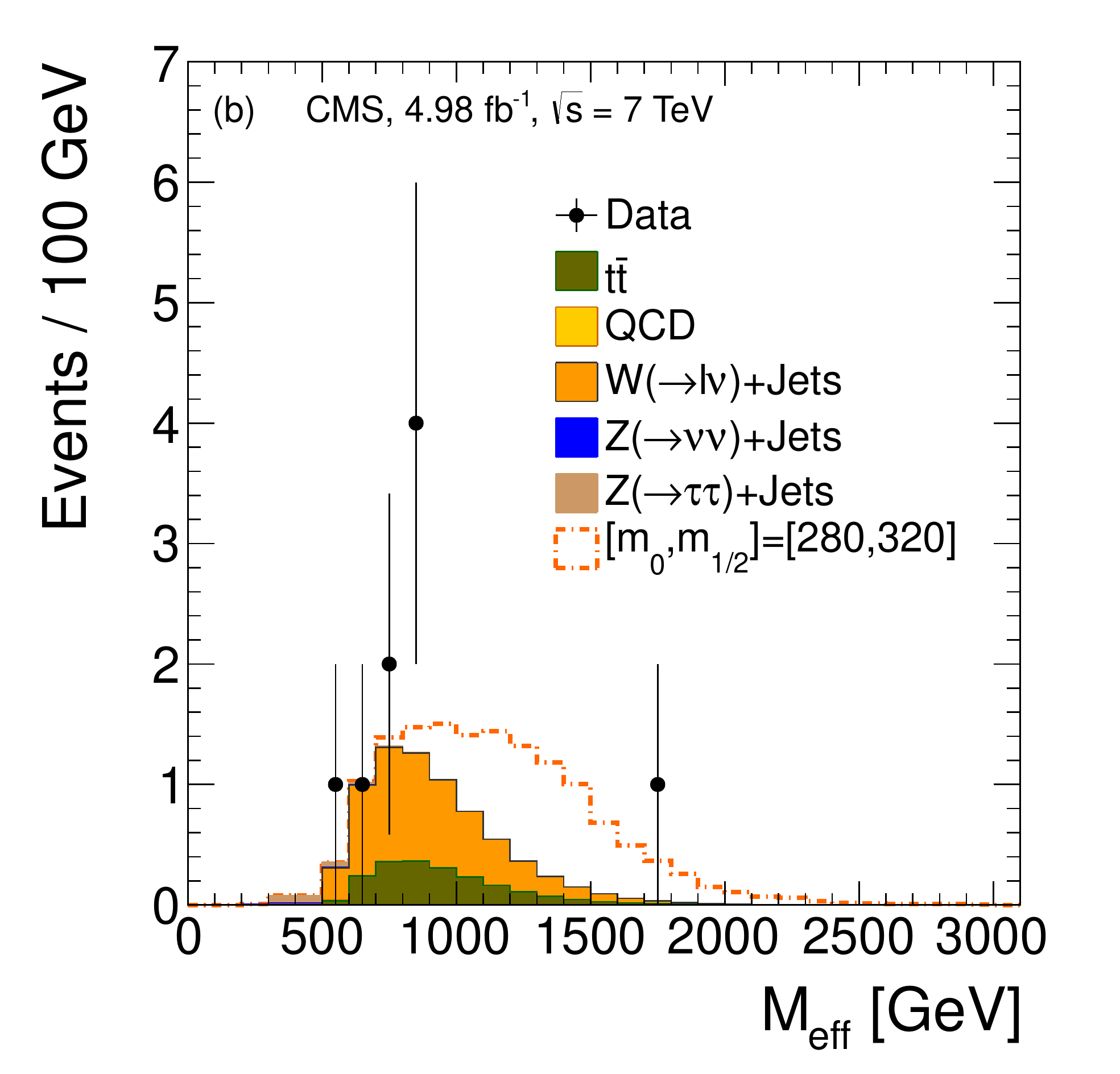}
\caption{Stacked distributions of (a) $\HT^{30}$, and (b) $M_\text{eff}$ in the SR for the multiple-\tauh\ final state. The background distributions are taken
from MC events that are normalized to the predictions based on data over the full region. The shapes obtained from MC simulation are used for illustrative
purposes only.} \label{fig:SignalPlot}
\end{center}
\end{figure}

The largest sources of background for the multiple-\tauh\
final state are from \ttbar and $\PW+\text{jets}$ events.
A counting experiment is performed and the background predictions from data are compared
with the observed number of events.
Table~\ref{table:expectations} lists these background predictions and the observed number of events in the SR.
Figure~\ref{fig:SignalPlot} shows the $\HT^{30}$ as well as the $M_\text{eff}$
distributions in the SR, where $M_\text{eff}$ is the sum $\MHT + \HT^{30}$.
The background distributions in Fig.~\ref{fig:SignalPlot} are taken
from simulation and normalized over the full spectrum.
The estimated number of events due to the SM background processes
is in agreement with the number of observed events in the SR.

\begin{table}[ht]
  \topcaption{Number of data and estimated background events with statistical and systematic uncertainties, respectively, in the multiple-\tauh\ final state.}
  \centering{
  \begin{tabular}{l  c c}
  \hline
       Process              & Signal Region          \\ [0.5ex] \hline
       QCD multijet events                  & $0.02 \pm 0.02  \pm 0.17$\\
       ${\rm W}+ $jets             & $5.20 \pm 0.63  \pm 0.62 $ \\
       $\ttbar$     & $2.03 \pm 0.36  \pm 0.34 $ \\
       $\cPZ(\to\tau\tau)+$ jets  & $0.21 \pm 0.13  \pm 0.17 $ \\
       $\cPZ(\to\nu\nu)+$ jets    & $0.03 \pm 0.02  \pm 0.50 $ \\
       Estimated $\sum SM$  & $7.49 \pm 0.74  \pm 0.90 $\\
  \hline
       Data    & 9 \\
  \hline

  \end{tabular}
  }
  \label{table:expectations}
\end{table}

\section{Limits on new physics}
\label{sec:limit}

\begin{figure}[htbp]
\begin{center}
\includegraphics[width=0.49\textwidth]{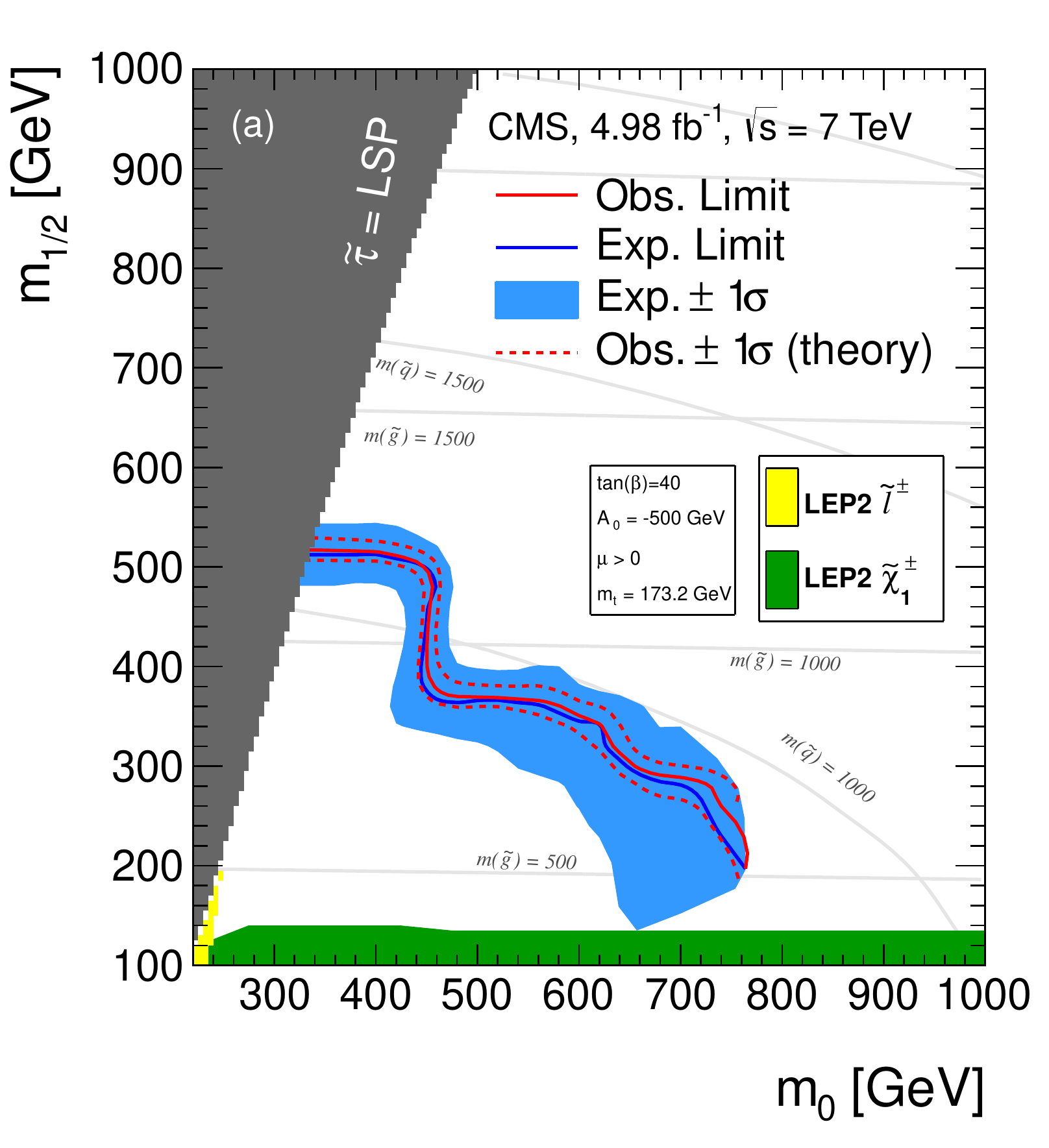}
\includegraphics[width=0.49\textwidth]{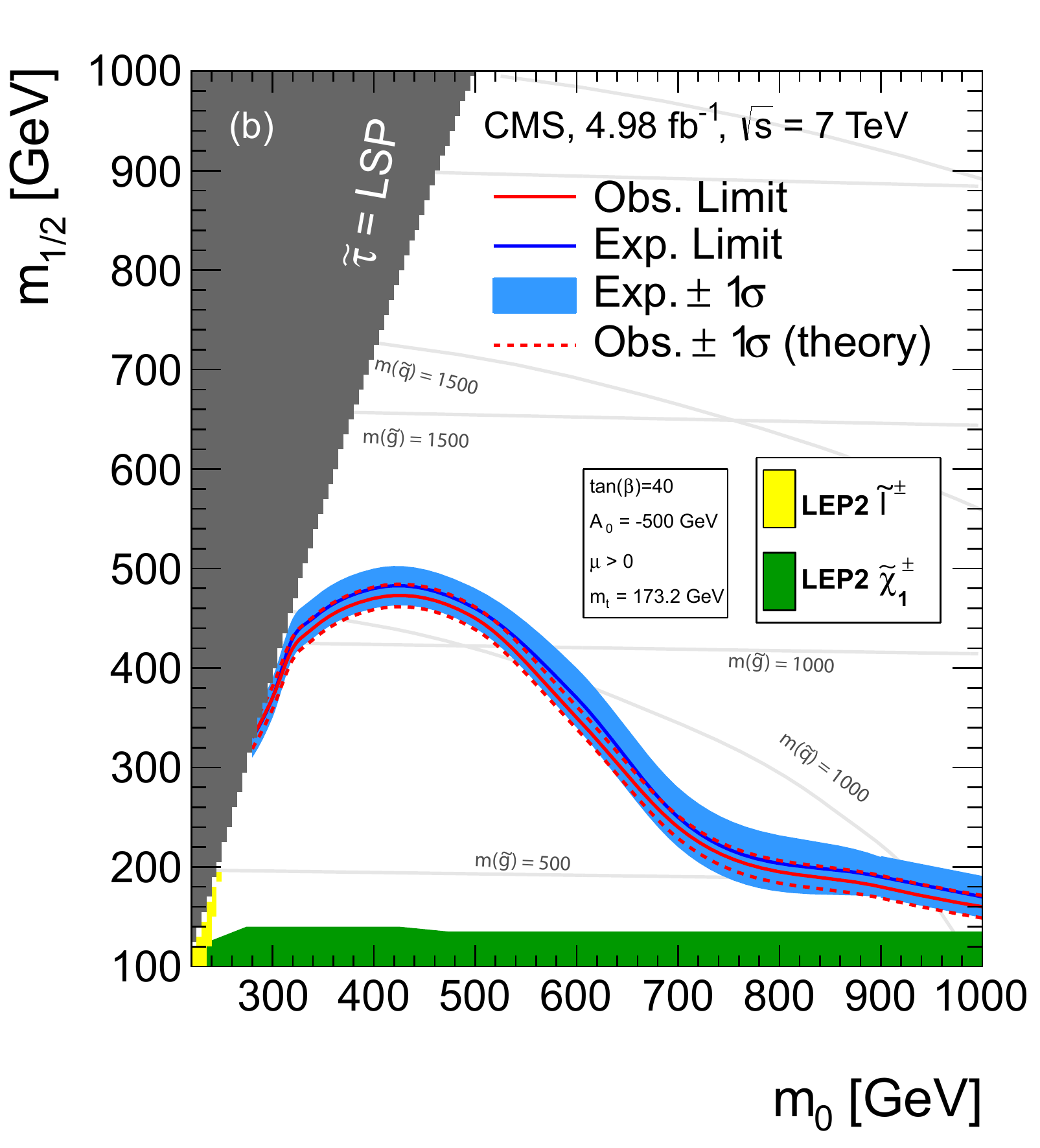}
\end{center}
\caption{95\% CL exclusion limits in the CMSSM plane at $\tan\beta=40$ for: (a) Single-\tauh final state, and (b) multiple-\tauh final state.
In the figures shown, the solid red line (Obs. Limit) denotes the experimental limit while the dotted red lines
(Obs. ${\pm}\sigma$ (theory) ) represent the uncertainty on the experimental limit due to uncertainties on the theoretical
cross sections. The blue band (Exp. ${\pm}\sigma$) represents the expected uncertainties. The contours of constant squark
and gluino mass are in units of \GeV.
}
\label{fig:Limit}
\end{figure}

The observed numbers of events in the single-$\tauh$ and multiple-$\tauh$ final states do not
reveal any evidence of physics beyond the standard model. Exclusion limits are set using the CL$_\mathrm{s}$ \cite{CLs}
criterion in the context of the CMSSM \cite{ref:CMSSMplot}.
The CMSSM parameter space with $\tan\beta=40$, $A_0= - 500$\GeV, $\mu>0$, and
$M_{\cPqt}=173.2$\GeV is chosen as a possible scenario with a light $\PSGt$
and a value of $\Delta M \le 20$\GeV. The excluded regions
are shown for the single-$\tauh$ and multiple-$\tauh$ final states in Figs. \ref{fig:Limit}(a) and \ref{fig:Limit}(b), respectively.
The limits are set using a simple counting experiment.
Systematic uncertainties are treated as nuisance parameters and marginalized, and
contamination from signal events in the control samples is taken into account.
In the CLs method, both the background-only as well as the signal $+$ background hypothesis are used to derive the confidence levels CL$_\mathrm{s}$ and the resulting limits and the uncertainty bands on the exclusion contours.
In the case of very small values of $\Delta M({\sim}5$\GeV), the lower-energy \tauh\
cannot be effectively detected and only the energetic \tauh\ from the decay of the
neutralino can be observed. The search for new physics with a single $\tau$ lepton has a
better sensitivity in this case. The single-\tauh\ and multiple-\tauh\ topologies thus have
complementary sensitivity and together provide coverage for models with a wide range of $\Delta M$ values.

Using the limits set by the single-\tauh\ analysis, a common gaugino mass $m_{1/2}$ of $<$495\GeV is excluded at 95\% Confidence Level (CL)
for a common scalar mass $m_{0}$ of $<$440\GeV.
For the multiple-\tauh\ analysis, $m_{1/2}< 465\GeV$ is excluded at 95\% CL for $m_{0}=440\GeV$.
A gluino with mass $<$1.15\TeV is excluded at 95\% CL for $m_0 < 440\GeV$. It can be noted that the
single-\tauh\ analysis shows better sensitivity for small values of $\Delta M$, which is near the boundary of $\PSGt=$ LSP.

\begin{figure}[htb]
  \centering
  {\includegraphics[width=.45\textwidth]{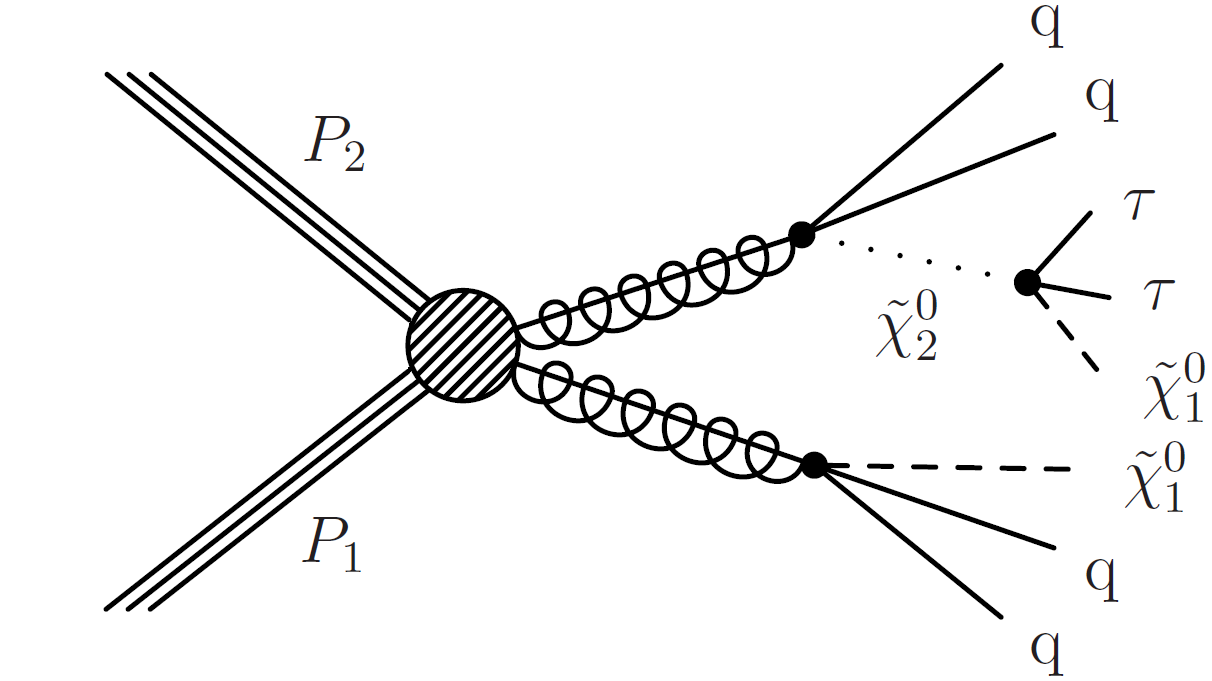}}
  \caption{Diagram for the T3tauh SMS model.}
  \label{fig:SMSdiagrams}
\end{figure}

\begin{figure}[htbp]
  \centering
\includegraphics[width=.49\textwidth]{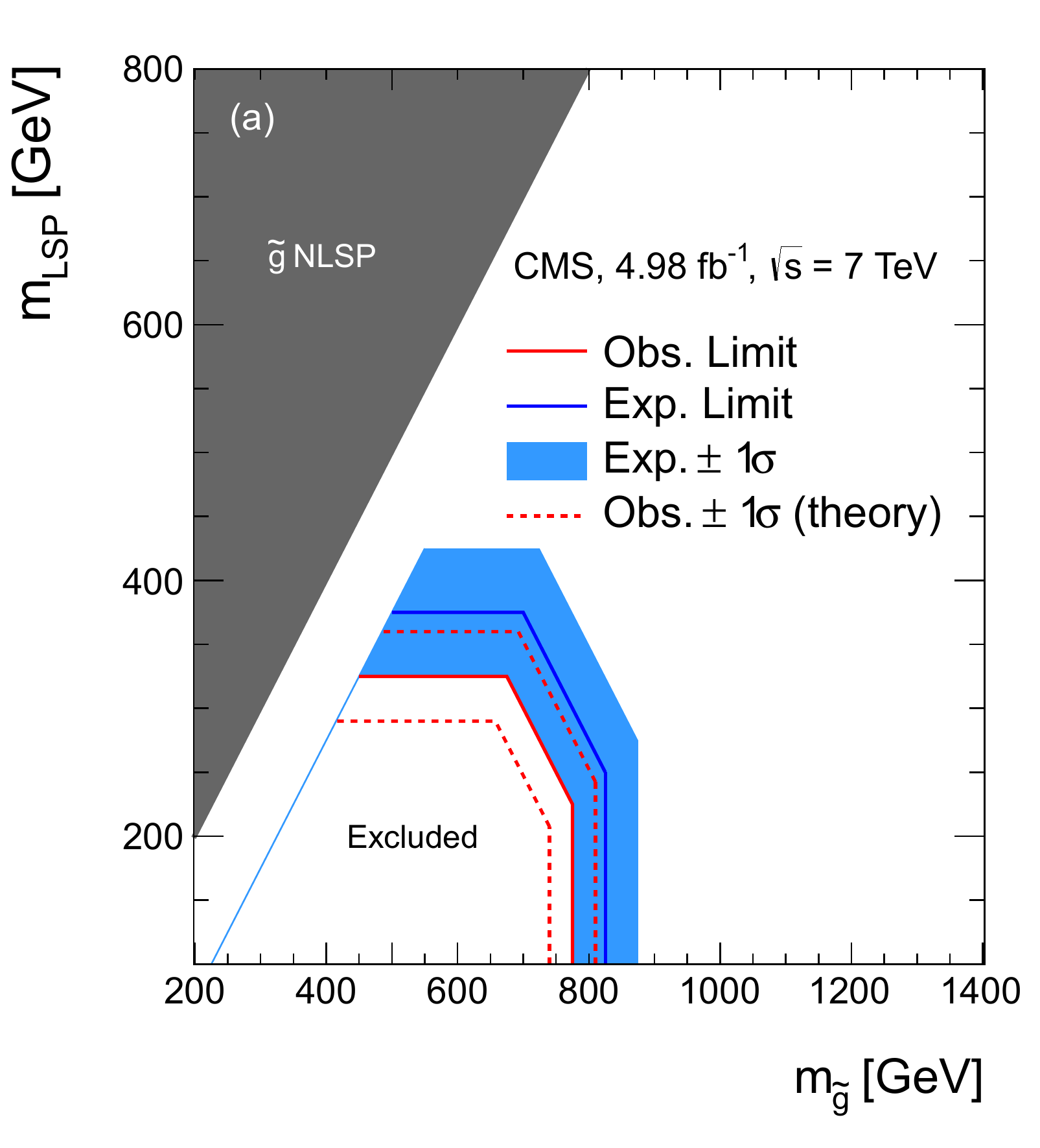}
\includegraphics[width=.49\textwidth]{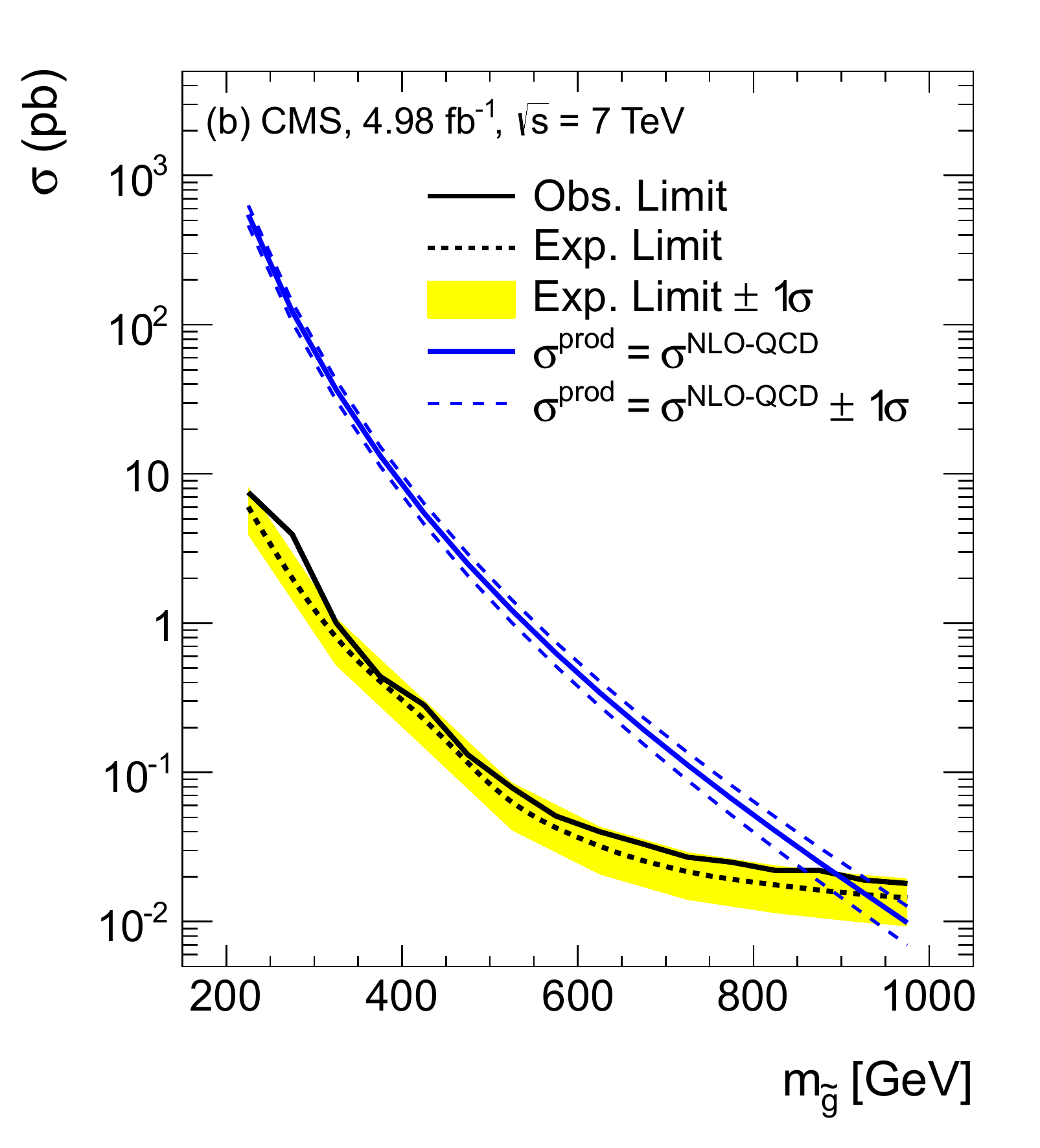}
\caption{Exclusion limits for the multiple-\tauh\ final state: (a) 95\% CL exclusion region obtained for the T3tauh model, where the solid red line
represents the limits on the mass of the gluino and the LSP; (b) 95\% CL cross section upper limits as a function of gluino mass in the GMSB scenario.
In this figure $\sigma^{prod}$ represents the cross section for the production of a pair of gluinos with subsequent decay into $\tau$ lepton pairs at
a 100\% branching fraction.
}
  \label{fig:SMSscans}
\end{figure}

The results for the multiple-\tauh\ final states are also interpreted in the context of
SMS \cite{SMS2}. The
$\tau\tau$ SMS scenario (T3tauh) is studied where
gluinos are produced in pairs and subsequently decay to $\tau$ lepton pairs and
an LSP via a neutralino
($\PSg \to \cPq\cPaq\PSGczDt$; $\PSGczDt \to \tau\overline{\tau} \to \tau\tau\PSGczDo$).
The diagram for the T3tauh model is given in Fig.~\ref{fig:SMSdiagrams}.
A gluino mass of $<$740\GeV is excluded at 95\% CL for LSP masses up to 205\GeV (here,
the mass of $\PSGczDt$ is the average of the masses of the gluino and the LSP).
Figure~\ref{fig:SMSscans}(a) shows the 95\% CL exclusion region obtained for T3tauh.
The limits on the mass of the gluino and LSP are shown with a solid red line.

In the simplified GMSB scenario,
the $\PSGt$ is the NLSP and decays to a
$\tau$ lepton and a gravitino $\PXXSG$, with a mass of the order of $\sim$\keV
\cite{GMSB1,GMSB2,GMSB3}
($\PSGczDt\to\tau\PSGt\to\tau\tau\PXXSG$). The topology for
this simplified GMSB scenario is similar to that of T3tauh except for the assumption that
both the gluinos decay to $\tau$-lepton pairs with a branching fraction of 100\%.
Therefore, the results are also interpreted in the simplified GMSB scenario using
the T3tauh scenario.
The signal acceptance is corrected to account for the final state containing up to
four $\tau$ leptons. A gluino with mass $<$860\GeV is excluded at 95\% CL.
Figure~\ref{fig:SMSscans}(b) shows the exclusion limits for the simplified GMSB scenario
as a function of the gluino mass.

Since the SMS topologies considered in this paper
are characterized by two $\tau$ leptons in the final state, we do not present SMS limits for the single-$\tauh$ final state.

\section{Summary}
\label{sec:conclusion}

A search for physics beyond the standard model with one or more hadronically decaying $\tau$ leptons, highly energetic jets, and large transverse momentum
imbalance in the final
state is presented.
The data sample corresponds to an integrated luminosity of $4.98 \pm 0.11\fbinv$ of pp
collisions at $\sqrt{s}=7$\TeV collected with the CMS detector.
The final number of events selected in data is consistent with the predictions
for standard model processes. We set upper limits on the cross sections for the CMSSM, GMSB, and SMS scenarios.
Within the CMSSM framework at $\tan\beta=40$, a gaugino mass $m_{1/2} <$ 495\GeV is excluded at
95\% CL for scalar masses $m_{0}<440\GeV$. This result sets a lower limit on the mass of the gluino at $1.15$\TeV with 95\% CL in this region.
In the multiple-$\tauh$ final state,
a gluino with a mass less than 740\GeV is excluded for the T3tauh simplified model
while a gluino with a mass less than 860\GeV is excluded for the simplified
GMSB scenario at 95\% CL.

\section*{Acknowledgements}
{\tolerance=800
\hyphenation{Bundes-ministerium Forschungs-gemeinschaft Forschungs-zentren}
We congratulate our colleagues in the CERN accelerator departments for the excellent performance of the LHC and thank the technical and
administrative staffs at CERN and at other CMS institutes for their contributions to the success of the CMS effort. In addition, we
gratefully acknowledge the computing centres and personnel of the Worldwide LHC Computing Grid for delivering so effectively the
computing infrastructure essential to our analyses. Finally, we acknowledge the enduring support for the construction and operation of
the LHC and the CMS detector provided by the following funding agencies:
the Austrian Federal Ministry of Science and Research; the Belgian Fonds de la Recherche Scientifique,
and Fonds voor Wetenschappelijk Onderzoek; the Brazilian Funding Agencies (CNPq, CAPES, FAPERJ, and FAPESP); the Bulgarian Ministry
of Education and Science; CERN; the Chinese Academy of Sciences, Ministry of Science and Technology, and National Natural Science
Foundation of China; the Colombian Funding Agency (COLCIENCIAS); the Croatian Ministry of Science, Education and Sport; the Research
Promotion Foundation, Cyprus; the Ministry of Education and Research, Recurrent financing contract SF0690030s09 and European Regional
Development Fund, Estonia; the Academy of Finland, Finnish Ministry of Education and Culture, and Helsinki Institute of Physics; the Institut National de Physique Nucl\'eaire et de Physique des Particules~/~CNRS, and Commissariat \`a l'\'Energie Atomique et aux \'Energies Alternatives~/~CEA, France; the Bundesministerium f\"ur Bildung und Forschung, Deutsche Forschungsgemeinschaft, and Helmholtz-Gemeinschaft Deutscher Forschungszentren, Germany; the General Secretariat for Research and Technology, Greece; the National Scientific Research Foundation, and National Office for Research and Technology, Hungary; the Department of Atomic Energy and the Department of Science and Technology, India; the Institute for Studies in Theoretical Physics and Mathematics, Iran; the Science Foundation, Ireland; the Istituto Nazionale di Fisica Nucleare, Italy; the Korean Ministry of Education, Science and Technology and the World Class University program of NRF, Korea; the Lithuanian Academy of Sciences; the Mexican Funding Agencies (CINVESTAV, CONACYT, SEP, and UASLP-FAI); the Ministry of Science and Innovation, New Zealand; the Pakistan Atomic Energy Commission; the Ministry of Science and Higher Education and the National Science Centre, Poland; the Funda\c{c}\~ao para a Ci\^encia e a Tecnologia, Portugal; JINR (Armenia, Belarus, Georgia, Ukraine, Uzbekistan); the Ministry of Education and Science of the Russian Federation, the Federal Agency of Atomic Energy of the Russian Federation, Russian Academy of Sciences, and the Russian Foundation for Basic Research; the Ministry of Science and Technological Development of Serbia; the Secretar\'{\i}a de Estado de Investigaci\'on, Desarrollo e Innovaci\'on and Programa Consolider-Ingenio 2010, Spain; the Swiss Funding Agencies (ETH Board, ETH Zurich, PSI, SNF, UniZH, Canton Zurich, and SER); the National Science Council, Taipei; the Scientific and Technical Research Council of Turkey, and Turkish Atomic Energy Authority; the Science and Technology Facilities Council, UK; the US Department of Energy, and the US National Science Foundation.

Individuals have received support from the Marie-Curie programme and the European Research Council (European Union); the Leventis Foundation; the A. P. Sloan Foundation; the Alexander von Humboldt Foundation; the Austrian Science Fund (FWF); the Belgian Federal Science Policy Office; the Fonds pour la Formation \`a la Recherche dans l'Industrie et dans l'Agriculture (FRIA-Belgium); the Agentschap voor Innovatie door Wetenschap en Technologie (IWT-Belgium); the Council of Science and Industrial Research, India; the Compagnia di San Paolo (Torino); and the HOMING PLUS programme of Foundation for Polish Science, cofinanced from European Union, Regional Development Fund.
\par}

\bibliography{auto_generated}   

\cleardoublepage \appendix\section{The CMS Collaboration \label{app:collab}}\begin{sloppypar}\hyphenpenalty=5000\widowpenalty=500\clubpenalty=5000\textbf{Yerevan Physics Institute,  Yerevan,  Armenia}\\*[0pt]
S.~Chatrchyan, V.~Khachatryan, A.M.~Sirunyan, A.~Tumasyan
\vskip\cmsinstskip
\textbf{Institut f\"{u}r Hochenergiephysik der OeAW,  Wien,  Austria}\\*[0pt]
W.~Adam, E.~Aguilo, T.~Bergauer, M.~Dragicevic, J.~Er\"{o}, C.~Fabjan\cmsAuthorMark{1}, M.~Friedl, R.~Fr\"{u}hwirth\cmsAuthorMark{1}, V.M.~Ghete, J.~Hammer, N.~H\"{o}rmann, J.~Hrubec, M.~Jeitler\cmsAuthorMark{1}, W.~Kiesenhofer, V.~Kn\"{u}nz, M.~Krammer\cmsAuthorMark{1}, I.~Kr\"{a}tschmer, D.~Liko, I.~Mikulec, M.~Pernicka$^{\textrm{\dag}}$, B.~Rahbaran, C.~Rohringer, H.~Rohringer, R.~Sch\"{o}fbeck, J.~Strauss, A.~Taurok, W.~Waltenberger, G.~Walzel, E.~Widl, C.-E.~Wulz\cmsAuthorMark{1}
% ...
\vskip\cmsinstskip
\dag:~Deceased\\
1:~~Also at Vienna University of Technology, Vienna, Austria\\
% ...

\end{sloppypar}
\end{document}